\documentclass[journal=nalefd,manuscript=letter]{achemso}

\usepackage[T1]{fontenc}
\usepackage{lmodern}
\usepackage{graphicx}
\usepackage{amsmath,amssymb}
\usepackage{siunitx}
\usepackage{booktabs}
\usepackage{microtype}
\usepackage{placeins}
\usepackage{url}
\usepackage{xcolor}
\usepackage{hyperref}
\hypersetup{hidelinks}

\newcommand{\kb}{k_{\mathrm{B}}}
\newcommand{\Tc}{T_{\mathrm{C}}}
\newcommand{\Th}{T_{\mathrm{H}}}
\newcommand{\q}{q}
\newcommand{\Eg}{E_{\mathrm{g}}}
\newcommand{\muEH}{\mu_{\mathrm{eh}}}
\newcommand{\kappaC}{\kappa}
\newcommand{\AM}{AM1.5G}
\newcommand{\dd}{\mathrm{d}}

\title{Fundamental Efficiency Limits of Transition-Metal Dichalcogenide Solar Cells with Carrier Multiplication and Hot-Carrier Effects}

\author{Seungwoo Lee}
\affiliation{Department of Integrated Energy Engineering (College of Engineering), KU-KIST Graduate School of Converging Science and Technology, and Department of Biomicrosystem Technology, Korea University, Seoul 02841, Republic of Korea}
\email{seungwoo@korea.ac.kr}


\begin{document}

\begin{abstract}
Detailed-balance limits (also called Shockley--Queisser (SQ) limits) for transition-metal dichalcogenide (TMD) solar cells have been reported, but existing TMD-specific limits do not simultaneously resolve thickness-dependent optics, carrier multiplication (CM), hot-carrier (HC) extraction, and finite cooling leakage.
Here, we develop a generalized detailed balance theory, defining a thermodynamically explicit upper-bound framework for this problem.
The model combines (i) energy- and thickness-dependent absorptance $a(E,d)$, including exciton-resolved monolayer absorbance, (ii) an experimentally available CM quantum-yield limit ($\eta_{\mathrm{CM}}\le0.97$), and (iii) an endoreversible HC engine with ideal energy-selective contacts and a finite heat-leak coefficient $\kappa$.
The framework includes an explicit resource-accounting proof showing that CM and HC draw on the same above-gap photon-energy reservoir: consequently, CM does not raise the reversible HC thermodynamic limit. Instead, CM can only provide finite-$\kappa$ protection by shifting part of the excess-energy utilization from a cooling-sensitive voltage channel into collected current.
For optically thick TMDs under AM1.5G illumination, the SQ optimum lies near $E_g\simeq1.3$~eV, whereas the CM/HC-favored envelope shifts toward $E_g\simeq1.0$~eV with reversible efficiencies above 50\%.
For monolayer TMDs (e.g., WSe$_2$ with $E_g=1.63$~eV), CM is essentially inactive because only $\sim3.7\%$ of above-gap AM1.5G photons satisfy $E>2E_g$, yielding an idealized short-circuit current gain of only $\sim0.6\%$ even before device nonidealities.
Bulk-like TMDs can show large HC-related gains at $d\sim10$--50~nm, but the cooling-budget analysis shows that even $\kappa=0.2~\mathrm{W\,m^{-2}\,K^{-1}}$ corresponds to a $\sim100~\mathrm{W\,m^{-2}}$ heat leak for $\Delta T=500$~K, i.e., an already highly aspirational cooling-suppression target.
The conclusion is therefore sharply constrained: for high-$E_g$ monolayer TMDs, CM is not a promising one-sun route, whereas narrow-$E_g$, bulk-like TMD absorbers remain plausible beyond-SQ candidates only if energy-selective extraction and phonon-engineered cooling suppression can be realized together.
\end{abstract}

\begin{tocentry}
\includegraphics[width=\linewidth]{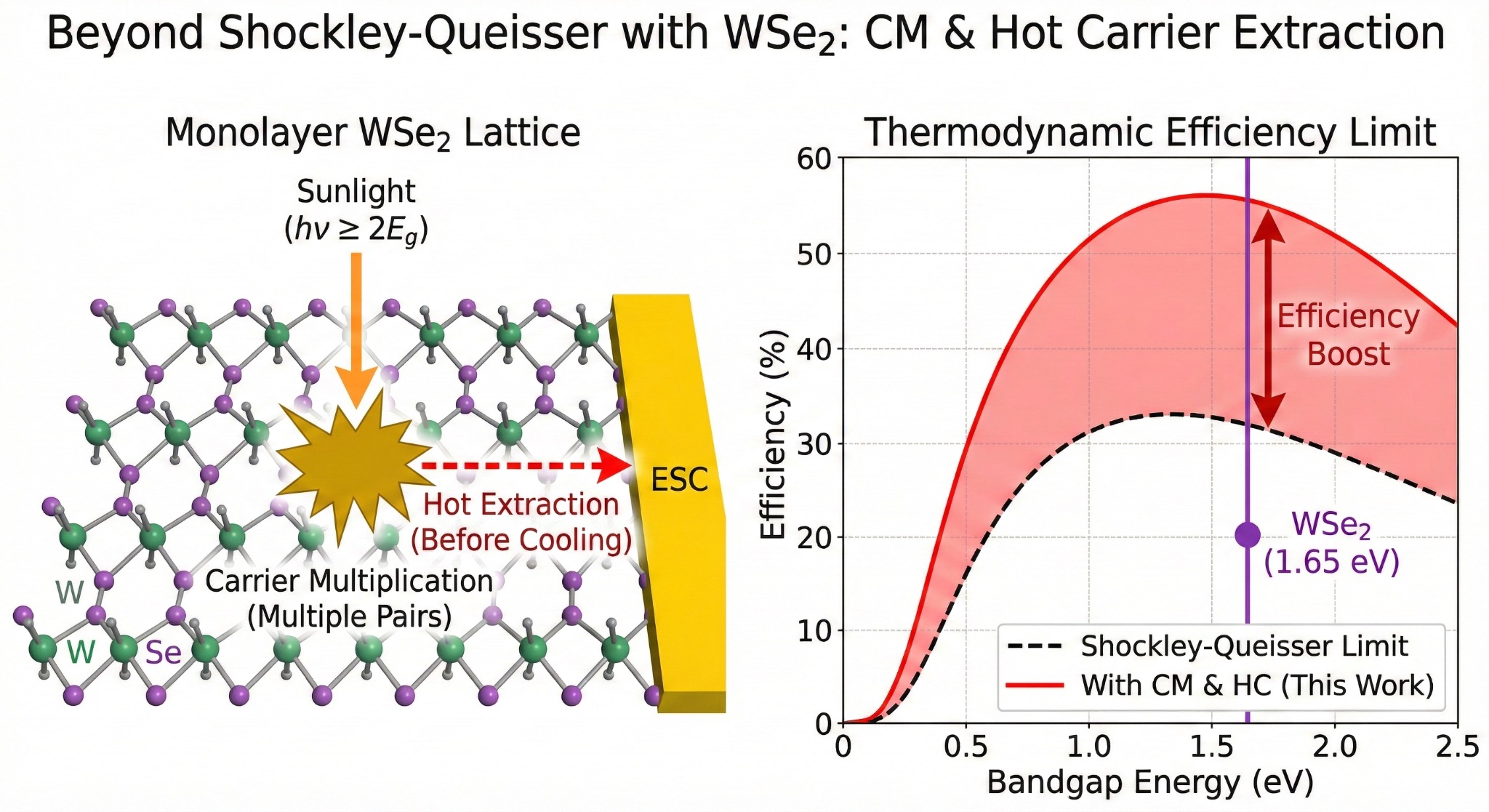}
\end{tocentry}

\section{Introduction}
Transition-metal dichalcogenides (TMDs) provide a distinctive photovoltaic testbed owing to their strong excitonic light--matter interaction, mechanical flexibility, and van der Waals (vdWs) heterointegration, which would be difficult to achieve with other platforms.\cite{Grossman2013NanoLett,Pop2023CommunPhys}
The conventional Shockley--Queisser (SQ) framework assumes an ideal step-function absorber and therefore obscures two features that are central for TMD devices: (i) absorptance depends strongly on thickness $(d)$ and optical stack, and (ii) the electronic bandgap $\Eg$ itself changes between monolayer and bulk-like forms for many TMDs.\cite{Pop2023CommunPhys,Fan2010LightTrapping,Miller2023Thickness}
Building on these realities, Pop et al. incorporated experimentally constrained, energy- and $d$-dependent absorptance $a(E,d)$ into detailed balance to obtain $d$-dependent SQ limits for TMD stacks.\cite{Pop2023CommunPhys}

Beyond SQ, TMDs also motivate interest in two nonequilibrium routes for reducing thermalization loss: carrier multiplication (CM) by inverse Auger/impact ionization and hot-carrier (HC) extraction enabled by energy-selective contacts (ESCs) in vdWs heterostructures.\cite{NatCommun2019CM,Werner1995APL,Queisser1996CM,Klimov2006APL,HannaNozik2006JAP,NatRevPhys2021HotCarrier,RossNozik1982}
CM is intrinsically thresholded by $2\Eg$ and therefore draws on a narrow, high-energy subset of the solar spectrum, whereas HC extraction is broadband in the sense that any above-$\Eg$ excess energy can, in principle, contribute to work if carriers are extracted before electron--phonon cooling.
The two mechanisms must, however, be compared with thermodynamic care: they are different conversion pathways for the same above-$\Eg$ photon-energy resource, not independent reservoirs of free energy.
Foundational thermodynamic formalisms for CM---from the observation by Werner, Brendel, and Queisser that the usual SQ assignment $\mu_\gamma=\q V$ must be modified for internal CM,\cite{Werner1995APL} to the thermodynamic limits of Brendel et al.,\cite{Queisser1996CM} and later detailed-balance CM analyses by Klimov\cite{Klimov2006APL} and by Hanna and Nozik\cite{HannaNozik2006JAP,DeVos1993,Sergeev2018}---and for HC extraction have, however, largely assumed idealized step absorbers or perfectly reversible cooling.
At the same time, TMD-focused detailed-balance studies have not yet unified these mechanisms with thickness-dependent and exciton-resolved TMD optics.
The resulting gap in the previous literature is not simply a missing efficiency number; it is a missing resource-accounting map that implies whether CM, HC extraction, absorber thickness, and the monolayer-to-bulk $\Eg$ crossover reinforce each other or merely repartition the same thermalization energy.
Therefore, it has remained unclear how experimentally constrained optics, monolayer-to-bulk $\Eg$ shifts, and finite cooling together shape the CM--HC interplay in TMD photovoltaics.

Here, we develop a generalized upper-bound detailed-balance limit for TMD solar cells that couples experimentally constrained optics, an idealized CM upper limit, and an endoreversible HC engine.
Specifically, we combine (i) $a(E,d)$ from optical constants for bulk films\cite{Pop2023CommunPhys} together with exciton-resolved monolayer absorbance spectra calibrated by a one-point A-exciton anchor,\cite{Grossman2013NanoLett,Pop2023CommunPhys} (ii) a CM quantum-yield upper bound motivated by experiment (up to $\eta_{\mathrm{CM}}=0.97$),\cite{NatCommun2019CM} and (iii) a De~Vos--Queisser-class HC model\cite{DeVos1993,Sergeev2018} with ESCs that optimizes the hot-reservoir temperature and chemical potential, augmented by a finite cooling-leakage coefficient $\kappa$ to quantify realizability.\cite{NatRevPhys2021HotCarrier}
We select WSe$_2$, MoS$_2$, and MoTe$_2$ as representative case studies, as they span the typical optical-$\Eg$ range of TMD semiconductors.
Because the model retains idealized ESCs and a lumped cooling-loss term, it should be interpreted as a thermodynamic upper-bound analysis rather than a predictive device simulation.
Within that scope, our unified CM--HC limit provides a quantitative beyond-SQ benchmark, linking spectral absorptance, $\Eg$, and electron--phonon cooling to the maximally achievable power-conversion efficiency of TMD photovoltaics. To make the resource accounting explicit, the framework contains three linked checks. First, the CM emission term is derived from the channel affinity of the inverse-Auger process rather than introduced as a free fitting ansatz. Second, the reversible $\kappa\rightarrow0$ degeneracy of HC and CM--HC is proven analytically and used as a thermodynamic consistency check. Third, the finite-$\kappa$ calculations are interpreted through a cooling-budget scale referenced to the one-sun power density, so that the numerical $\kappa$ values can be judged as upper-bound, aspirational, or strongly lossy regimes.

\begin{figure*}[!t]
\centering
\includegraphics[width=0.7\textwidth]{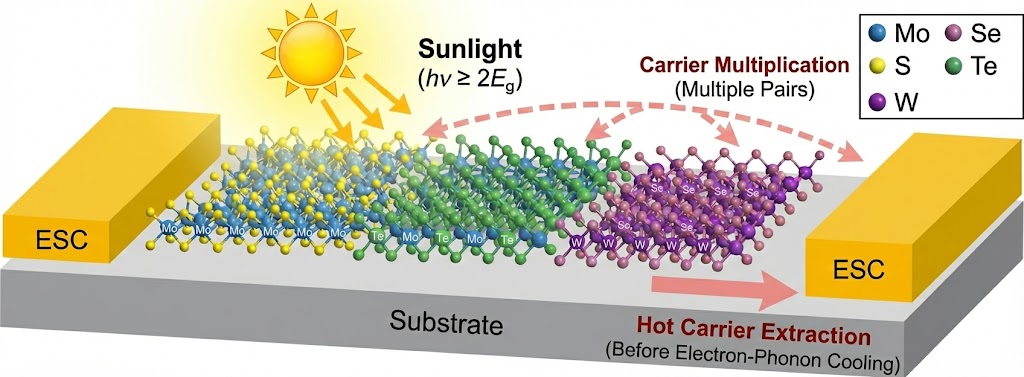}
\caption{Concept of carrier multiplication (CM) and hot-carrier (HC) interplay in TMDs.
CM is activated only by photons with $E\ge2\Eg$, whereas ESC-enabled HC extraction can, in principle, harvest thermalization power from the broader above-$\Eg$ spectrum before electron--phonon cooling.
Because both mechanisms draw on the same excess photon energy, CM does not raise the reversible HC ceiling; instead, at finite cooling leakage (parametrized by $\kappa$) it can redistribute part of that excess energy into current before it is lost.}
\label{fig:Concept}
\end{figure*}
\FloatBarrier

\subsection{Key theory}

\textbf{Figure~\ref{fig:Concept}} summarizes the coupled CM--HC pathway and motivates the unified formulation below.
The model summary and key parameter conventions are summarized in Section~S1 and Tables~S1--S3 of the Supporting Information (SI), while complete derivations are given in Sections~S2--S4 of the SI.
We denote photon energy by $E$ and the bandgap by $\Eg$, and use the hemispherical generalized black-body photon flux (Section~S1.1 of the SI):
\begin{equation}
\Phi_{\mathrm{bb}}(E,T,\mu_\gamma)=\frac{2\pi}{h^{3}c^{2}}\,
\frac{E^{2}}{\exp\!\left[(E-\mu_\gamma)/(\kb T)\right]-1}
\end{equation}
where $\mu_\gamma$, $c$, and $\kb$ are the photon chemical potential, the speed of light, and the Boltzmann constant, respectively.
Under \AM\ illumination (ASTM G-173-03), the absorbed solar photon flux is $a(E,d)\Phi_{\odot}(E)$ and the radiative emission is $a(E,d)\Phi_{\mathrm{bb}}$.\cite{ShockleyQueisser1961}
Section~S1.2 of the SI details the TMD-specific $a(E,d)$ used in this work.
The SQ current is therefore
\begin{equation}
J_{\mathrm{SQ}}(V)=\q\!\int_{0}^{\infty}\! a(E,d)\Big[\Phi_{\odot}(E)-\Phi_{\mathrm{bb}}(E,\Tc,\mu_\gamma=\q V)\Big]\,dE
\label{eq:JSQ}
\end{equation}
where $\Tc$ is the cell (or lattice) temperature.
Equation~\ref{eq:JSQ} reduces to the standard SQ limit for $a(E,d)=\Theta(E-\Eg)$.

CM modifies radiative detailed balance, because an emitted photon can annihilate more than one electron--hole pair.\cite{Werner1995APL,Queisser1996CM}
This is not an ad hoc nonequilibrium ansatz. In the early CM detailed-balance literature, Werner, Brendel, and Queisser explicitly pointed out that for internal CM, the usual SQ assignment $\mu_\gamma=\q V$ is invalid and that the radiative saturation current must be modified accordingly;\cite{Werner1995APL} Brendel et al. then used that generalized radiative balance to derive CM thermodynamic limits,\cite{Queisser1996CM} and later CM efficiency-limit studies continued to use detailed-balance formalisms of this class.\cite{Klimov2006APL,HannaNozik2006JAP}
In that sense, the energy dependence belongs to the \emph{radiative channel stoichiometry}, not to multiple electronic quasi-Fermi-level splittings. The carrier reservoir is still characterized by a single pair chemical potential ($\mu_{\mathrm{eh}}$ in the HC formulation or $\q V$ in the isothermal CM limit); the factor $m(E)$, corresponding to an energy-dependent number of collected electron--hole pairs per absorbed
photon, appears, because the inverse radiative channel at energy $E$ annihilates $m(E)$ pairs.
As established in thermodynamic treatments of inverse Auger processes, emission channels that consume $m(E)$ pairs must therefore carry the elevated photon chemical potential $\mu_\gamma=m(E)\q V$ to preserve macroscopic entropy balance within an endoreversible upper-bound closure.\cite{Werner1995APL,Queisser1996CM,Klimov2006APL,HannaNozik2006JAP}
In a mean-multiplicity formulation (the SI discusses channel-resolved generalizations), we write
\begin{equation}
J_{\mathrm{CM}}(V)=\q\!\int_{0}^{\infty}\! a(E,d)\,m(E)\Big[\Phi_{\odot}(E)-\Phi_{\mathrm{bb}}(E,\Tc,\mu_\gamma=m(E)\q V)\Big]\,dE
\label{eq:JCM}
\end{equation}
where $m(E)=1$ for $E<2\Eg$ and $m(E)=1+\eta_{\mathrm{CM}}\left(\frac{E}{\Eg}-2\right)$ for $E>2\Eg$ (Section~S1.3 of the SI).
Here, we use $\eta_{\mathrm{CM}}=0.97$ as an optimistic experimental upper bound from ultrafast spectroscopy.\cite{NatCommun2019CM}
Operational photovoltaic yields may be lower, so the CM results in our work should be interpreted as limit values rather than as expected device performance.

The quantity that is energy-dependent is the photon chemical potential that is assigned to a radiative channel with multiplicity $m(E)$; the electronic reservoir itself is still described by one pair chemical potential ($qV$ in the isothermal CM limit or $\mu_{\mathrm{eh}}$ in the HC formulation). Thus, $\mu_\gamma(E)=m(E)qV$ expresses the stoichiometry of the inverse radiative channel, not multiple energy-dependent electronic quasi-Fermi splittings. This is the same macroscopic detailed-balance closure used in the previously reported CM efficiency-limit literature.\cite{Werner1995APL,Queisser1996CM,Klimov2006APL,HannaNozik2006JAP}

HC extraction is described by a HC reservoir with temperature $\Th$ and chemical potential $\muEH$, coupled to cold contacts at $\Tc$ through idealized ESCs with energy separation $\Delta\varepsilon$ (Sections~S3--S4 of the SI).
The particle balance resembles Eq.~\ref{eq:JSQ} but with $\Tc\to\Th$ and $\mu_\gamma\to\muEH$ (and with $m(E)$ included if CM is also present).
For reversible extraction, ESCs impose the De~Vos relation\cite{DeVos1993,Sergeev2018}
\begin{equation}
\q V=\frac{\Tc}{\Th}\muEH+\left(1-\frac{\Tc}{\Th}\right)\Delta\varepsilon
\label{eq:devos}
\end{equation}
which indicates that the chemical potential of the photon and thus the resulting efficiency can be further improved by the HC-driven energy separation $\Delta\varepsilon$, weighted by $1-\frac{\Tc}{\Th}$. If $\Th=\Tc$ (HC is not present), Eq.~\ref{eq:devos} reduces to $\q V=\muEH$, as in the normal SQ limit for the voltage. As such, the present HC treatment is an endoreversible upper-bound model: it assumes ideal ESCs and captures departures from reversibility only through a lumped cooling-leak term.

To represent finite phonon-mediated cooling (or any parasitic heat flow), we add a phenomenological phonon heat-leak term
$\dot Q_{\mathrm{cool}}=\kappaC(\Th-\Tc)$
in the hot-reservoir energy balance, which removes energy from the HC population. In other words, $\dot Q_{\mathrm{cool}}$ reduces $\Delta\varepsilon$ and the resulting voltage/output power $J(V)V$ over $(V,\muEH,\Th,\Delta\varepsilon)$, as will be further detailed below (also see the details in Section~3--4 of SI). This produces the HC-only and CM--HC results in our work.

Interpreting $\kappaC$ as an effective area-normalized cooling conductance, we use $\kappaC=0$ as the reversible upper bound and $\kappaC=0.2~\mathrm{W\,m^{-2}\,K^{-1}}$ as an aspirational finite-leak benchmark.
To place this value in context, one-sun \AM\ illumination carries $\sim1000~\mathrm{W\,m^{-2}}$ of incident power.
For an elevated HC temperature difference $\Delta T\equiv\Th-\Tc\approx500$~K, $\kappaC=0.2$ corresponds to $\dot Q_{\mathrm{cool}}\approx100~\mathrm{W\,m^{-2}}$ (about $10~\mathrm{mW\,cm^{-2}}$, or roughly 10\% of the incident power).
Thus, $\kappaC=0.2$ should be read as an optimistic regime requiring strongly suppressed cooling rather than as a generic experimental value.\cite{NatRevPhys2021HotCarrier}

\subsection{TMD-specific optics inputs}

For bulk TMD films, we adopt the optical-constant-based absorptance $a(E,d)$ derived from the $4n^{2}$ light-trapping bound and measured optical constants (Section~S1.2 of the SI).\cite{Pop2023CommunPhys,Fan2010LightTrapping}
For monolayers, absorptance spectra are calibrated to a one-point A-exciton anchor and, when optical constants are reported as absorption coefficients, mapped to single-pass absorptance for a 1~nm thickness; complete definitions and anchors are provided in Section~S1.2 and Tables~S1--S3 of the SI.\cite{Pop2023CommunPhys,Grossman2013NanoLett}
Because the detailed-balance efficiencies scale with the energy-resolved absorbed-photon flux, this anchoring fixes the absolute absorptance scale and prevents arbitrary rescaling of $\eta$.
A critical subtlety is the crossover of $\Eg$ itself.
For example, monolayer WSe$_2$ has a direct $\Eg$ near $\sim1.6$~eV, whereas its bulk counterpart becomes indirect near $\sim1.3$~eV.\cite{Lee2020NanoLettContact,Pop2023CommunPhys}
Therefore, we treat monolayer and bulk-like TMD limits, and we emphasize throughout that WSe$_2$, MoS$_2$, and MoTe$_2$ are representative case studies rather than an exhaustive surrogate for all TMD chemistries.

\section{Bandgap sets whether CM or HC can matter}

\textbf{Figure~2(a)} summarizes the $\Eg$-dependent upper limits for an optically thick absorber under \AM\ illumination.
The bulk gaps of MoTe$_2$ ($\Eg$ of 1.04~eV), MoS$_2$ ($\Eg$ of 1.22~eV), and WSe$_2$ ($\Eg$ of 1.29~eV) are indicated by vertical gray lines.
The SQ efficiency peaks near $\sim30\%$ at $E_g\simeq1.3$~eV, placing bulk WSe$_2$ close to the conventional optimum.
CM shifts the optimum toward smaller $\Eg$, because a non-negligible fraction of absorbed photons must satisfy $E\gtrsim2\Eg$ for multiplication to matter.
Within the $\Eg$ window relevant to common bulk TMDs ($\sim1.0$--2.1~eV), this immediately favors MoTe$_2$-like $\Eg$ over larger-$\Eg$ WSe$_2$- or MoS$_2$-like systems.
The corresponding efficiency sensitivity with respect to $\Eg$ of MoTe$_2$ is detailed in Figure~S1 and Table~S4 of the SI (Section~S5.1).

\begin{figure*}
\centering
\includegraphics[width=\textwidth]{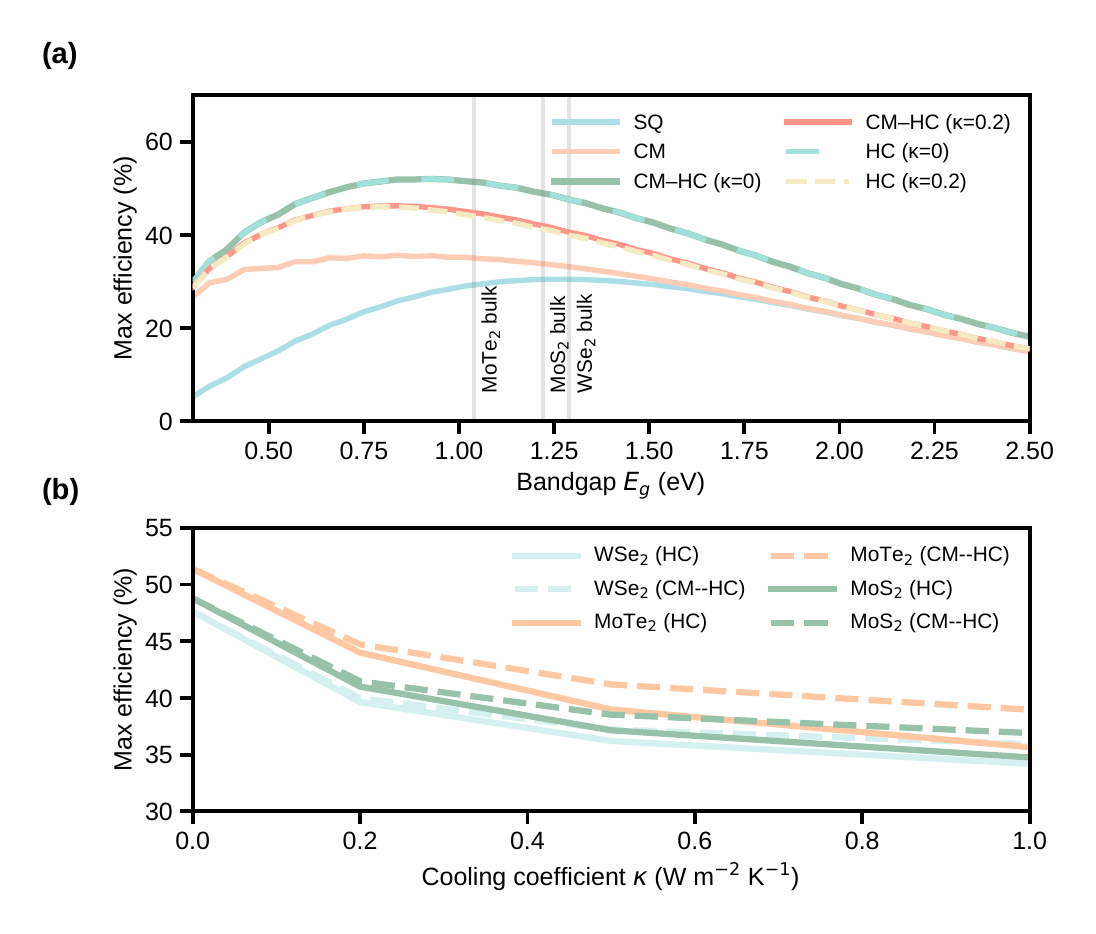}
\caption{$\Eg$- and cooling-leakage dependence of the detailed-balance limits under \AM\ illumination.
(a) Optically thick absorber ($a(E)=1$) with SQ, CM, HC, and CM--HC limits.
For readability, HC is plotted as dashed curves for $\kappaC=0$ and $0.2~\mathrm{W\,m^{-2}\,K^{-1}}$, while CM--HC is plotted as solid curves at the same $\kappaC$ values.
In the reversible limit ($\kappaC=0$), HC and CM--HC coincide because the optimum drives $\mu_{\mathrm{eh}}\to0$ (Sections~S3.1--S3.3 of the SI).
Vertical markers indicate representative bulk TMD bandgaps.
(b) $\kappaC$ sweep of the maximum efficiency for HC and CM--HC at the bulk $\Eg$ of WSe$_2$, MoTe$_2$, and MoS$_2$.
Together, (a,b) show that finite cooling leakage rapidly erodes HC-related efficiency gains, whereas CM primarily shifts the favored $\Eg$ toward $\sim1$~eV and modestly buffers finite-cooling losses.}
\label{fig:EgSweep}
\end{figure*}

In the reversible limit ($\kappaC=0$), the HC/CM--HC upper bound can exceed 50\% near $E_g\simeq1.0$~eV, because ESC-enabled extraction can convert part of the excess carrier enthalpy into voltage.
Given the absorbed and emitted energy flux densities,
\begin{align}
P_{\mathrm{abs}} &= \int_0^\infty E\,a(E)\phi_{\odot}(E)\,\dd E\\
P_{\mathrm{emit}}(\muEH,\Th) &= \int_0^\infty E\,a(E)\phi_{\mathrm{bb}}(E,\Th,\muEH)\,\dd E
\end{align}
the net radiative energy deposited into the HC absorber becomes $P_{\mathrm{abs}}-P_{\mathrm{emit}}$.
In the combined CM--HC case, the same radiative energy balance is evaluated with the energy-dependent photon chemical potential $\mu_\gamma(E)=m(E)\muEH$ in the emission term. If $\dot N = J/\q$ is the extracted carrier-pair flux, we define the average extracted carrier energy per pair $\Delta E$ via energy conservation:
\begin{equation}
\Delta E \equiv \frac{P_{\mathrm{abs}}-P_{\mathrm{emit}}-\dot Q_{\mathrm{cool}}}{\dot N}
\label{eq:deltaE_def}
\end{equation}
This $\Delta E$ is the energy window that the ESCs must transmit (electron extraction at $E_e$ and hole extraction at $E_h$ such that $\Delta E=E_e-E_h$).
In the endoreversible limit, $\Delta E$ enters directly into Eq.~\eqref{eq:devos}; at finite $\kappaC$, cooling subtracts from this window and therefore reduces the HC voltage and the extracted power.

The HC result at $\kappaC=0$ lies directly on top of the CM--HC curve at $\kappaC=0$.
This overlap is not a plotting artifact; it is a direct consequence of thermodynamics.
CM and HC are not independent degrees of freedom: both draw on the same excess photon energy above the $\Eg$.
In the reversible optimum, $\muEH\to 0$ because any finite $\muEH$ increases the emitted photon flux and therefore reduces the net extractable power.
Setting $\muEH=0$ in Eq.~\eqref{eq:devos} yields
\begin{equation}
\q V = \left(1-\frac{\Tc}{\Th}\right)\Delta E.
\label{eq:qV_mu0}
\end{equation}
Multiplying Eq.~\eqref{eq:qV_mu0} by $J$ and invoking Eq.~\eqref{eq:deltaE_def} gives the compact output-power expression
\begin{equation}
P = JV = \left(1-\frac{\Tc}{\Th}\right)\left(P_{\mathrm{abs}}-P_{\mathrm{emit}}-\kappaC(\Th-\Tc)\right).
\label{eq:P_mu0}
\end{equation}
For $\kappaC=0$, the right-hand side depends only on the net radiative power balance $P_{\mathrm{abs}}-P_{\mathrm{emit}}$.
CM changes the extracted particle flux through $m(E)$ but does not change the absorbed photon-energy flux $P_{\mathrm{abs}}$.
Moreover, when $\muEH=0$, the CM--HC emission spectrum collapses to the same $\mu_\gamma(E)=0$ black-body form as HC-only emission.
Therefore, HC and CM--HC share the same reversible limit: CM simply trades higher current for proportionally lower voltage at fixed extracted power. Also, it is noteworthy that we used the reversible degeneracy as a diagnostic rather than a weakness of the model: if CM--HC exceeded HC at $\kappa=0$ under otherwise identical optical inputs, the calculation would double count excess photon energy. The SI gives the corresponding channel-affinity derivation (see details in Section S3.5 and S3.6 of the SI) and a numerical check showing zero HC/CM--HC efficiency difference at $\kappa=0$ for the representative bulk TMD $\Eg$ (Section S5.2--S5.3 and Table S5--S6 of the SI), while finite positive differences appear only when the explicit heat-leak term is nonzero (Section S5.4 and Table S7 of the SI).

For $\kappaC>0$, such scaling invariance breaks down.
Because cooling penalizes high $\Th$ and degrades the HC voltage, CM can act as a kinetic buffer by converting a fraction of high-energy photons into additional electron--hole pairs before that excess energy is lost through the heat-leak channel.
The finite CM--HC enhancement over HC at nonzero $\kappaC$ therefore reflects redistribution of the same thermodynamic resource under irreversible cooling, not access to a new reversible efficiency limit.
Additional photovoltaic metrics ($J_\mathrm{sc}$, $V_\mathrm{oc}$, and fill factor where defined) for the representative TMD absorbers in \textbf{Figure~2(a)} are summarized in Table~S8 of the SI (Section~S5.5).

\textbf{Figure~2(b)} isolates the role of electron--phonon cooling by sweeping $\kappa$ at fixed bulk TMD $E_g$; the corresponding numerical values are summarized in Table~S9 of the SI (Section~S5.6).
As $\kappa$ increases, the cooling-leak term $\kappa(\Th-\Tc)$ forces a lower optimal HC temperature and a smaller electrochemical splitting at the ESCs, directly suppressing the HC voltage.
In contrast, the CM--HC limit retains the largely $\kappa$-insensitive current enhancement from CM of above-threshold photons ($E>2\Eg$), so the CM--HC efficiency decreases more slowly with $\kappa$ than the HC-only limit.
This is the sense in which CM ``helps'' at finite cooling: it partially shifts excess-energy utilization from a heat-leak-sensitive voltage channel into collected current.

\section{Bulk TMDs: thickness-dependent limits and cooling sensitivity}

\begin{figure*}
\centering
\includegraphics[width=\textwidth]{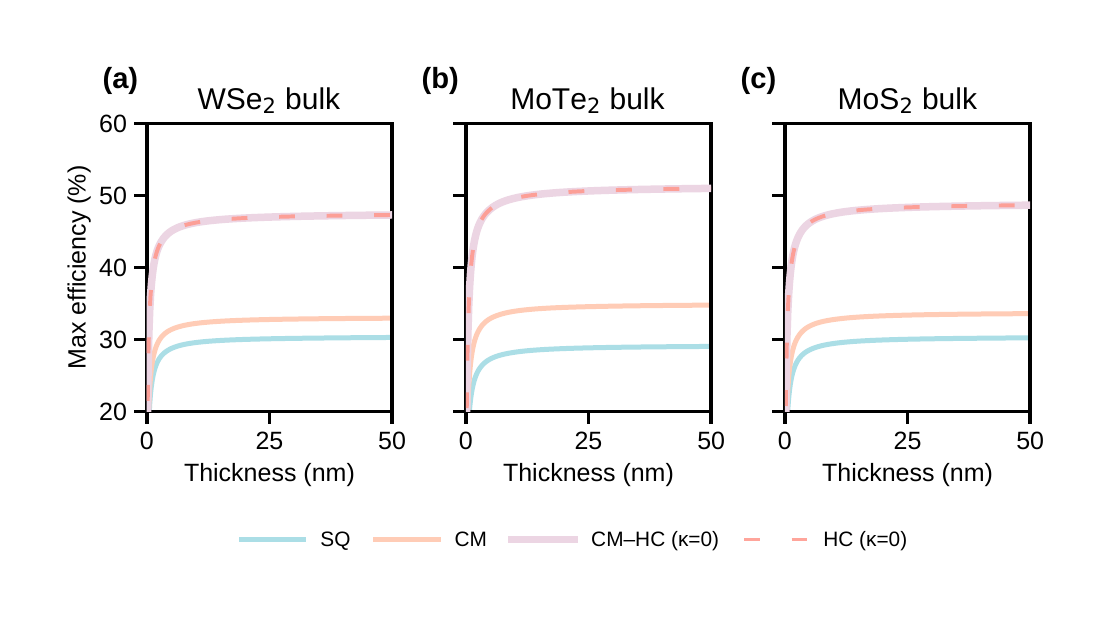}
\caption{Thickness ($d$)-dependent bulk TMD efficiency limits from optical-constant-based absorptance $a(E,d)$ (Yablonovitch light-trapping form with $n=4$).
Panels show WSe$_2$ ($E_g=1.29$~eV), MoTe$_2$ ($E_g=1.04$~eV), and MoS$_2$ ($E_g=1.22$~eV) over 0--50~nm.
Each panel overlays SQ, CM ($\eta_{\mathrm{CM}}=0.97$), HC, and CM--HC in the reversible limit ($\kappaC=0$).
In that reversible limit, HC and CM--HC coincide closely; the HC curve is drawn as a dashed overlay so the degeneracy remains visible.
The key message is that once films become optically strong (tens of nm and above), efficiency saturates with $d$ and is controlled primarily by $\Eg$ and cooling rather than by further thickness increases.}
\label{fig:Thickness}
\end{figure*}

\textbf{Figure~\ref{fig:Thickness}} applies the unified model to bulk-like WSe$_2$, MoTe$_2$, and MoS$_2$ using optical-constant-based $a(E,d)$.
These three materials are representative case studies chosen to span a typical bulk TMD $\Eg$ range from about 1.0 to 1.3~eV.
As $d$ increases, the absorptance envelope saturates and the corresponding SQ and beyond-SQ limits converge rapidly.
Thus, once films reach the tens-of-nanometers regime, further increases in $d$ provide diminishing thermodynamic returns and the dominant control variables become $\Eg$ and cooling rather than thickness itself.\cite{Fan2010LightTrapping,Miller2023Thickness}

For bulk WSe$_2$ ($\Eg=1.29$~eV) and MoS$_2$ ($\Eg=1.22$~eV), CM yields only a modest increase over SQ, because the fraction of \AM\ photons above $2\Eg$ is limited but non-negligible.
Reversible HC extraction ($\kappa=0$) yields a much larger efficiency gain, since conventional operation still discards substantial thermalization power.
However, that HC advantage is extremely sensitive to cooling.
The contrast between $\kappaC=0$ and $\kappaC=0.2~\si{\watt\per\meter\squared\per\kelvin}$ collapses the HC-gain window precisely in the thickness range where absorptance has nearly saturated.
For $\Delta T\approx500$~K, $\kappaC=0.2$ implies $\dot Q_{\mathrm{cool}}\approx100~\si{\watt\per\meter\squared}$, i.e., about 10\% of one-sun input power, highlighting that this is already a highly optimistic cooling-loss budget.\cite{NatRevPhys2021HotCarrier}
The $\kappa$ sensitivity underlying this feasibility challenge is quantified directly by the sweep in \textbf{Figure~2(b)} and Table~S9 of the SI.

For bulk MoTe$_2$ ($\Eg=1.04$~eV), the CM threshold $2\Eg\simeq2.08$~eV lies deeper in the visible, so CM is intrinsically more effective under \AM.
Simultaneously, the smaller $\Eg$ offers more thermalization energy that HC extraction can, in principle, harvest.
Thus, MoTe$_2$ sits near the CM/HC-favored $\Eg$ window of \textbf{Figure~\ref{fig:EgSweep}}, and its $d$-dependent limits show the largest beyond-SQ gains among the three representative cases, again strongly tempered by $\kappaC$ (\textbf{Figure~\ref{fig:Thickness}}).

\section{Monolayers: the optical transparency constraint and why CM can be negligible}

\begin{figure*}
\centering
\includegraphics[width=\textwidth]{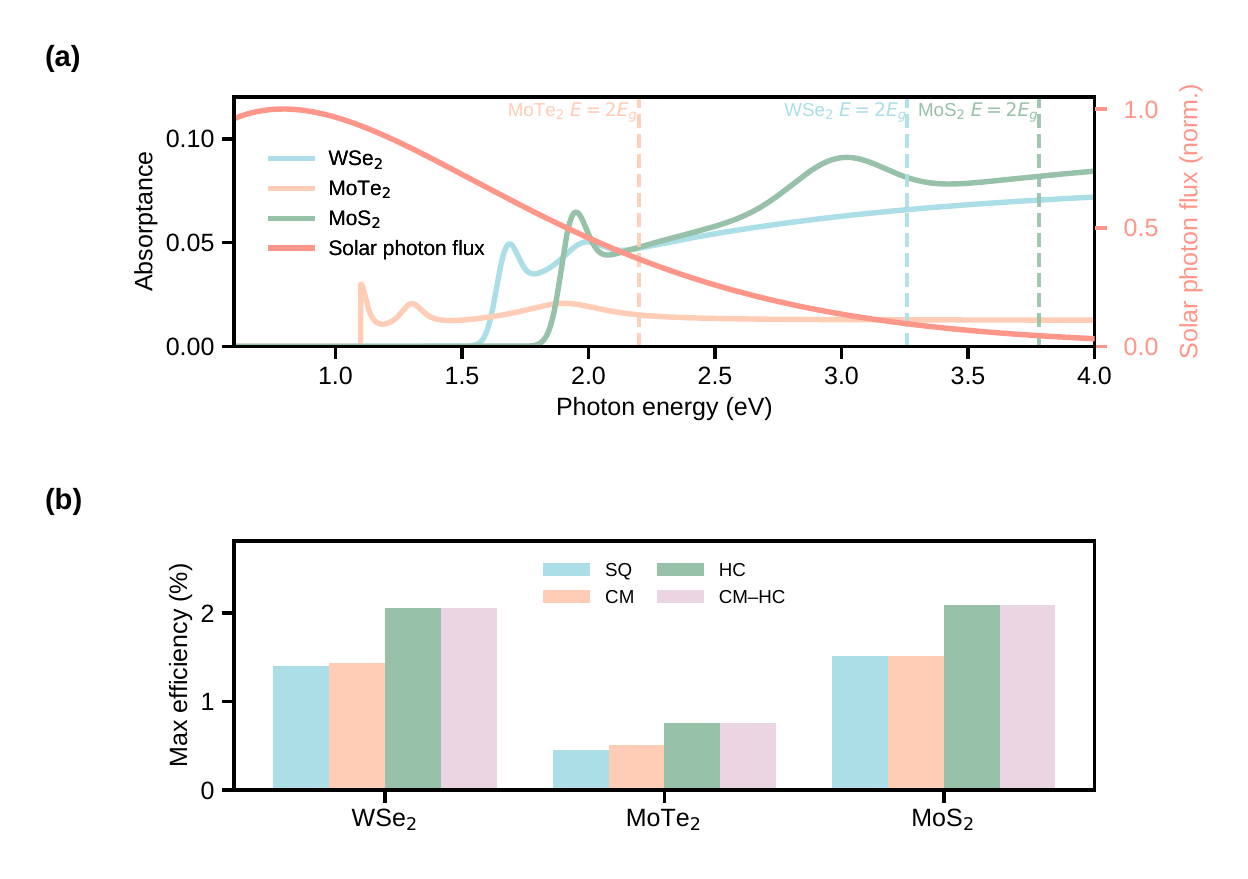}
\caption{TMD monolayer limits and the optical-transparency constraint.
(a) Absorptance spectra for WSe$_2$, MoTe$_2$, and MoS$_2$ monolayers (left axis) overlaid with the normalized solar photon flux (right axis).
Vertical dashed lines mark the $2\Eg$ thresholds for each monolayer.
(b) Maximum efficiencies for each monolayer, including SQ, CM, HC, and CM--HC (shown for $\kappaC=0$).
For large-gap monolayers such as WSe$_2$ and MoS$_2$, CM is intrinsically disfavored because the $2\Eg$ threshold lies deep in the UV (small available photon fraction), whereas MoTe$_2$-like gaps lower the threshold but remain constrained by thin-film absorptance.
Data sources and robustness checks are summarized in Section~S1.2 and Sections~S4--S5 of the SI.}
\label{fig:MonoTMD}
\end{figure*}

\textbf{Figure~\ref{fig:MonoTMD}} compares absorptance spectra (\textbf{Figure~\ref{fig:MonoTMD}(a)}) and achievable efficiencies (\textbf{Figure~\ref{fig:MonoTMD}(b)}) for three representative TMD monolayers (all taken as $t=1$~nm): WSe$_2$ ($\Eg\approx1.63~\mathrm{eV}$), MoS$_2$ ($\Eg\approx1.89~\mathrm{eV}$), and MoTe$_2$ ($\Eg\approx1.10~\mathrm{eV}$).
WSe$_2$ and MoS$_2$ absorptance spectra are taken from literature and tabulated in Section~S1.2 (Table~S1) of the SI.\cite{Pop2023CommunPhys,Grossman2013NanoLett}
MoTe$_2$ uses a compact excitonic absorptance model (background continuum plus Lorentzian A/B resonances anchored to the reported $\Eg$) with parameters in Table~S2 of the SI and cutoff-robustness validation in Table~S3 and Section~S1.2 of the SI.\cite{Ruppert2014MoTe2}
In all cases, the SQ efficiencies are only a few percent, because most incident photons are not absorbed.

CM is often negligible for a more fundamental reason: the solar spectrum contains too few photons above $2\Eg$ to generate meaningful extra pairs.
Quantitatively, for WSe$_2$, the fraction of above-$2\Eg$ photons is $f_{>2E_g}\approx3.7\%$ (\textbf{Figure~\ref{fig:MonoTMD}(a)}).
Those photons have a mean normalized excess energy $\langle E/E_g-2\rangle_{E>2E_g}\approx0.16$, so a Beard-type linear-yield model implies
$\Delta J_{\mathrm{sc}}/J_{\mathrm{sc}}\approx\eta_{\mathrm{CM}}\,f_{>2E_g}\langle E/E_g-2\rangle_{E>2E_g}\approx0.6\%$
even for $\eta_{\mathrm{CM}}=0.97$.
This is an upper bound on current gain before the radiative penalty in Eq.~\ref{eq:JCM} is included, and it is reduced further because monolayer absorbance is concentrated near excitonic resonances close to $\Eg$.
The same argument applies even more strongly to MoS$_2$.
MoTe$_2$ lowers the $2\Eg$ threshold into the visible/near-IR, but its overall performance remains constrained by thin-film absorptance unless optical enhancement strategies increase $a(E)$ across the solar band (\textbf{Figure~\ref{fig:MonoTMD}(b)}).
The corresponding $J_\mathrm{sc}$, $V_\mathrm{oc}$, and FF values are summarized in Table~S8 of the SI (Section~S5.5).
The SI further includes a CM stress-test table comparing SQ and CM efficiencies and current gains at the deliberately optimistic $\eta_{\mathrm{CM}}=0.97$ (Table~S10 of the SI (Section~S5.7)); this table shows that the monolayer conclusion remains unchanged even under the most favorable CM assumption used in the paper.

Finally, the comparison emphasizes that even if HC extraction is thermodynamically favorable in the reversible limit, monolayer performance is typically dominated by optics: without broadband absorption enhancement, the available absorbed power (and thus the thermalization loss that HC could recycle) remains small.
This is why pursuing CM in high-$\Eg$ monolayer TMDs under one-sun illumination is fundamentally unrewarding.

\section{Conclusion}
The unified model yields three central outcomes that guide TMD design.
First, in the optically thick limit under \AM\ illumination, the combined CM--HC upper bound shifts the favored bandgap from the SQ value near $E_g\simeq1.3$~eV toward $E_g\simeq1.0$~eV, identifying MoTe$_2$-like gaps as intrinsically favorable.
Second, monolayer WSe$_2$ is effectively CM-insensitive because only a small fraction of absorbed photons satisfy $E>2\Eg$ and the absorption itself is concentrated in near-edge excitonic resonances.
Third, bulk-like TMDs can exhibit large HC-related efficiency gains at modest $d$, but these gains collapse rapidly with increasing $\kappa$, making electron--phonon cooling the dominant feasibility constraint.

The calculations yield compact design rules for TMD photovoltaics beyond SQ.
(i) $\Eg$ is the primary lever: under one-sun \AM, the CM/HC-favored envelope lies near $\Eg\sim1$~eV, making MoTe$_2$-like gaps intrinsically favorable, whereas larger-gap monolayers are CM-disfavored because their $2\Eg$ thresholds sit in the UV.
(ii) For monolayers, absorption is the gating constraint; CM and HC cannot deliver meaningful efficiency gains until optical strategies raise $a(E)$ across the solar band.
(iii) For bulk-like TMD films that absorb strongly at $d\sim10$--50~nm, HC gains can be substantial, but they survive only if the effective cooling conductance remains well below order-unity \si{\watt\per\meter\squared\per\kelvin}.
(iv) From an experimental perspective, the most plausible direction for exceeding SQ limits is therefore not high-gap monolayer CM, but rather narrow-gap bulk-like TMD absorbers integrated with advanced ESCs and phonon-engineered transport layers that suppress carrier cooling.
Future work aimed at predictive device performance will need explicit nonequilibrium transport, Auger kinetics, and contact selectivity beyond the present endoreversible upper-bound framework; such modeling would refine the device-level interpretation, but it does not alter the value of the present thermodynamic benchmark.

\section*{Supporting Information}
SI is available free of charge.
It contains: (i) absorptance models and monolayer calibration anchors (Section~S1); (ii) the CM quantum-yield model $m(E)$ and its parameters (Section~S2); (iii) full SQ/CM detailed-balance derivations together with the HC/CM--HC energy balance, an expanded thermodynamic-consistency derivation, a channel-affinity proof for $\mu_\gamma(E)=m(E)\mu_{\mathrm{eh}}$, and physical grounding of the cooling-leakage term (Section~S3); (iv) a microscopic flux derivation of ESCs and the terminal-voltage relation (Section~S4); and (v) additional numerical checks, including MoTe$_2$ bandgap sensitivity, $\kappa$ sweeps, cooling-budget tables, CM-yield stress tests, and maximum-power current--voltage resource accounting (Section~S5).

\section*{Code and Data Availability}
Code, processed spectra, and numerical datasets used in the main text and SI are available in \href{https://github.com/NEOlab-code/Fundamental-Limits-to-Photovoltaics-with-Carrier-Multiplication-and-Hot-Carrier-Effects}{the public GitHub repository}: https://github.com/NEOlab-code/Fundamental-Limits-to-Photovoltaics-with-Carrier-Multiplication-and-Hot-Carrier-Effects. The repository includes scripts that generate the thermodynamic checks reported here: the reversible HC/CM--HC degeneracy test, the finite-$\kappa$ cooling-budget table, the CM upper-bound stress test, and the current--voltage resource-accounting tables reported in the SI.

\section*{Author Information}
\subsection*{Corresponding Author}
Seungwoo Lee --- E-mail: seungwoo@korea.ac.kr

\subsection*{Author Contributions}
S.L. conceived the original research idea, computed all results, and wrote the manuscript.

\subsection*{Notes}
The author declares no competing financial interest.

\section*{Acknowledgments}
We acknowledge funding from National Research Foundation of Korea (RS-2022-NR068141) and from the KIST Institutional Program (Project No.: 2V09840-23-P023).
This research was also supported by a grant of the Korea--US Collaborative Research Fund (KUCRF), funded by the Ministry of Science and ICT and Ministry of Health \& Welfare, Republic of Korea (grant number: RS-2024-00468463), and by Korea University grant.



\begin{thebibliography}{99}

\bibitem{Grossman2013NanoLett}
Bernardi, M.; Palummo, M.; Grossman, J. C. Extraordinary Sunlight Absorption and One Nanometer Thick Photovoltaics Using Two-Dimensional Monolayer Materials. \textit{Nano Lett.} \textbf{2013}, \textit{13}, 3664--3670. DOI: 10.1021/nl401544y.

\bibitem{Pop2023CommunPhys}
Nassiri Nazif, K.; Nitta, F. U.; Daus, A.; Saraswat, K. C.; Pop, E. Efficiency Limit of Transition Metal Dichalcogenide Solar Cells. \textit{Commun. Phys.} \textbf{2023}, \textit{6}, 367. DOI: 10.1038/s42005-023-01447-y.

\bibitem{Fan2010LightTrapping}
Yu, Z.; Raman, A.; Fan, S. Fundamental Limit of Nanophotonic Light Trapping in Solar Cells. \textit{Proc. Natl. Acad. Sci. U.S.A.} \textbf{2010}, \textit{107}, 17491--17496. DOI: 10.1073/pnas.1008296107.

\bibitem{Miller2023Thickness}
Miller, D. A. B. Why Optics Needs Thickness. \textit{Science} \textbf{2023}, \textit{379}, 41--45. DOI: 10.1126/science.ade3395.

\bibitem{NatCommun2019CM}
Kim, J. H.; Bergren, M. R.; Park, J. C.; Adhikari, S.; Lorke, M.; Frauenheim, T.; Choe, D.-H.; Kim, B.; Choi, H.; Gregorkiewicz, T.; Lee, Y. H. Carrier Multiplication in van der Waals Layered Transition Metal Dichalcogenides. \textit{Nat. Commun.} \textbf{2019}, \textit{10}, 5488. DOI: 10.1038/s41467-019-13325-9.

\bibitem{Werner1995APL}
Werner, J. H.; Brendel, R.; Queisser, H. J. Radiative Efficiency Limit of Terrestrial Solar Cells with Internal Carrier Multiplication. \textit{Appl. Phys. Lett.} \textbf{1995}, \textit{67}, 1028--1030. DOI: 10.1063/1.114719.

\bibitem{Queisser1996CM}
Brendel, R.; Werner, J. H.; Queisser, H. J. Thermodynamic Efficiency Limits for Semiconductor Solar Cells with Carrier Multiplication. \textit{Sol. Energy Mater. Sol. Cells} \textbf{1996}, \textit{41--42}, 419--425. DOI: 10.1016/0927-0248(95)00125-5.

\bibitem{Klimov2006APL}
Klimov, V. I. Detailed-Balance Power Conversion Limits of Nanocrystal-Quantum-Dot Solar Cells in the Presence of Carrier Multiplication. \textit{Appl. Phys. Lett.} \textbf{2006}, \textit{89}, 123118. DOI: 10.1063/1.2356314.

\bibitem{HannaNozik2006JAP}
Hanna, M. C.; Nozik, A. J. Solar Conversion Efficiency of Photovoltaic and Photoelectrolysis Cells with Carrier Multiplication Absorbers. \textit{J. Appl. Phys.} \textbf{2006}, \textit{100}, 074510. DOI: 10.1063/1.2356795.

\bibitem{NatRevPhys2021HotCarrier}
Paul, K. K.; Kim, J.-H.; Lee, Y. H. Hot Carrier Photovoltaics in van der Waals Heterostructures. \textit{Nat. Rev. Phys.} \textbf{2021}, \textit{3}, 178--192. DOI: 10.1038/s42254-020-00272-4.

\bibitem{RossNozik1982}
Ross, R. T.; Nozik, A. J. Efficiency of Hot-Carrier Solar Energy Converters. \textit{J. Appl. Phys.} \textbf{1982}, \textit{53}, 3813--3818. DOI: 10.1063/1.331124.

\bibitem{DeVos1993}
De Vos, A. The Endoreversible Theory of Solar Energy Conversion: A Tutorial. \textit{Sol. Energy Mater. Sol. Cells} \textbf{1993}, \textit{31}, 75--93. DOI: 10.1016/0927-0248(93)90008-Q.

\bibitem{Sergeev2018}
Sergeev, A.; Sablon, K. Exact Solution, Endoreversible Thermodynamics, and Kinetics of the Generalized Shockley--Queisser Model. \textit{Phys. Rev. Applied} \textbf{2018}, \textit{10}, 064001. DOI: 10.1103/PhysRevApplied.10.064001.

\bibitem{ShockleyQueisser1961}
Shockley, W.; Queisser, H. J. Detailed Balance Limit of Efficiency of p-n Junction Solar Cells. \textit{J. Appl. Phys.} \textbf{1961}, \textit{32}, 510--519. DOI: 10.1063/1.1736034.

\bibitem{Lee2020NanoLettContact}
Yang, S.; Cha, J.; Kim, J. C.; Lee, D.; Huh, W.; Kim, Y.; Lee, S. W.; Park, H. G.; Jeong, H. Y.; Hong, S.; Lee, G. H.; Lee, C. H. Monolithic Interface Contact Engineering to Boost Optoelectronic Performances of 2D Semiconductor Photovoltaic Heterojunctions. \textit{Nano Lett.} \textbf{2020}, \textit{20}, 2443--2451. DOI: 10.1021/acs.nanolett.9b05162.

\bibitem{Ruppert2014MoTe2}
Ruppert, C.; Aslan, O. B.; Heinz, T. F. Optical Properties and Band Gap of Single- and Few-Layer MoTe$_2$ Crystals. \textit{Nano Lett.} \textbf{2014}, \textit{14}, 6231--6236. DOI: 10.1021/nl502557g.

\end{thebibliography}
\end{document}


\maketitle

\section{S1. Model summary and parameter conventions}
\subsection{S1.1 Spectral variables and photon fluxes}
We work in photon energy $E=\hbar\omega$ and use the incident \AM\ solar photon flux spectrum $\phi_{\odot}(E)$ (photons\,m$^{-2}$\,s$^{-1}$\,eV$^{-1}$).
The radiative emission of a planar photovoltaic device is written in terms of the generalized Planck spectrum (photon chemical potential $\mu_\gamma$)
\begin{equation}
\phi_{\mathrm{bb}}(E,T,\mu_\gamma) \equiv \frac{2E^2}{h^3c^2}\frac{1}{\exp\!\left(\frac{E-\mu_\gamma}{\kb T}\right)-1}
\label{eq:planck}
\end{equation}
where the prefactor corresponds to emission per unit area per unit energy into a hemisphere for a Lambertian emitter (the geometric factors and refractive-index $n$ corrections used by Pop et al.\cite{Pop2023CommunPhys} are summarized below). In all detailed balance (DB) calculations, the cell temperature is set by $\Tc=300$~K.
\subsection{S1.2 Absorptance models}
\paragraph{Bulk-like films (optical-constant-based $a(E,d)$).}
For thick or multilayer TMD films, we follow Pop et al.\cite{Pop2023CommunPhys} and describe absorption via an energy $(E)$- and thickness $(d)$-dependent absorptance $a(E,d)$ rather than the ideal step-function $a=\Theta(E-\Eg)$.
The central optical input of Pop et al.\cite{Pop2023CommunPhys} can be interpreted as a smooth, DB-compatible upper bound that combines (i) measured optical constants and (ii) the classical Lambertian light-trapping bound.

\paragraph{From the $4n^{2}$ (Tiedje--Yablonovitch) light-trapping bound to a closed-form $a(E,d)$.}
In the ray-optics limit with perfect angular randomization and a perfect rear reflector, the path-length enhancement factor is bounded by\cite{Yablonovitch1982,Tiedje1984,Fan2010LightTrapping}
\begin{equation}
F \le 4n^{2}
\end{equation}
where $n$ is the refractive index and the $4n^{2}$ factor originates from the small escape cone of an isotropic photon gas inside a high-$n$ medium.
Equivalently, the probability that a randomized internal ray falls within the external escape cone scales as $p_{\mathrm{esc}}\sim 1/(4n^{2})$, so the mean number of traversals before escape scales as $N\sim 1/p_{\mathrm{esc}}\sim 4n^{2}$.
This motivates an effective mean optical path length
\begin{equation}
L_{\mathrm{eff}}\sim F d \sim 4n^{2}d
\label{eq:Leff}
\end{equation}
For a photon ``trapped'' in this randomized slab, absorption and escape are competing loss channels.
If absorption occurs with rate $\alpha(E)$ per unit length and escape occurs with an effective rate $1/L_{\mathrm{eff}}$, then the probability that the photon is absorbed before it escapes is the branching ratio
\begin{equation}
a(E,d)=\frac{\alpha(E)}{\alpha(E)+\frac{1}{L_{\mathrm{eff}}}}
=\frac{\alpha(E)L_{\mathrm{eff}}}{1+\alpha(E)L_{\mathrm{eff}}}
\label{eq:a_branch}
\end{equation}
Substituting Eq.~\eqref{eq:Leff} yields the closed form used by Pop et al.:\cite{Pop2023CommunPhys}
\begin{equation}
a(E,d)=\frac{4n^{2}\alpha(E)d}{1+4n^{2}\alpha(E)d}
=\frac{\alpha(E)}{\alpha(E)+\frac{1}{4n^{2}d}}
\label{eq:pop_absorptance}
\end{equation}
Eq.~\eqref{eq:pop_absorptance} has the correct limiting behaviors:
for weak absorption ($\alpha d\ll 1/(4n^{2})$), $a\approx 4n^{2}\alpha d$ (the maximum enhancement over a single-pass $\alpha d$),
whereas for strong absorption ($\alpha d\gg 1/(4n^{2})$), $a\to 1$ (saturation).
Pop et al.\cite{Pop2023CommunPhys} further discuss parasitic absorption in transport layers; in that case, one may replace $\alpha\to \alpha+\alpha_{\mathrm{par}}$ in the denominator (SI of Ref.~\citenum{Pop2023CommunPhys}).
We stress that Eq.~\eqref{eq:pop_absorptance} is a ray-optics, angle-averaged bound.
For high-$n$ thin films (tens of nm), interference can introduce oscillations about this envelope, and recent bounds emphasize that broadband optical performance requires finite thickness.\cite{Miller2023Thickness}
In our work, we use Eq.~\eqref{eq:pop_absorptance} as a physically transparent upper bound consistent with the DB treatment of Pop et al.,\cite{Pop2023CommunPhys} and we report efficiency as $\eta(d)$ rather than a single ``optically thick'' number.
\paragraph{Monolayer WSe$_2$ and MoS$_2$.}
We calibrate each digitized monolayer absorptance spectrum by one-point A-exciton anchoring,
\begin{equation}
a(E)=s\,a_{\mathrm{shape}}(E),\qquad
s=\frac{a_{\mathrm{ref}}(E_A)}{a_{\mathrm{shape}}(E_A)}
\label{eq:mono_anchor}
\end{equation}
so that $a(E_A)=a_{\mathrm{ref}}(E_A)$ (Table~\ref{tab:mono_abs_anchor}), while preserving the spectral line shape.
For WSe$_2$ and MoS$_2$, Pop et al.\cite{Pop2023CommunPhys} report absorption coefficients $\alpha(E)$; we convert these to absorptance for thickness $d=1$~nm via Beer--Lambert law,
\begin{equation}
a(E)=1-\exp[-\alpha(E)d]
\label{eq:alpha_to_a}
\end{equation}
before applying Eq.~\eqref{eq:mono_anchor}. The calibrated absorption spectra used in \textbf{Figure~4a} of Main text are available in \href{https://github.com/NEOlab-code/Fundamental-Limits-to-Photovoltaics-with-Carrier-Multiplication-and-Hot-Carrier-Effects}{the public GitHub repository}.

\begin{table}[H]
\centering
\caption{Monolayer absorptance calibration anchors used to set the absolute scale of the digitized spectra.
Each spectrum is multiplied by a constant factor so that $a(E_A)=a_{\mathrm{ref}}(E_A)$ (one-point calibration), while preserving the spectral shape.}
\label{tab:mono_abs_anchor}
\centering
\begin{tabular}{@{}lll p{0.6\textwidth}@{}}
\hline
Material & $E_A$ (eV) & $a_{\mathrm{ref}}(E_A)$ & Reference / note \\
\hline
WSe$_2$ (mono) & 1.69 & 0.049 & A-exciton anchor from Ref.~\cite{Pop2023CommunPhys} (SI Fig.~1; $\alpha$ converted via Eq.~\eqref{eq:alpha_to_a} with $d=1$~nm) \\
MoS$_2$ (mono) & 1.975 & 0.061 & A-exciton peak absorbance from Ref.~\cite{Grossman2013NanoLett} (Fig.~1c) \\
MoTe$_2$ (mono) & 1.10 & 0.026 & Free-standing absorbance shown in Ref.~\cite{Ruppert2014MoTe2}(Fig.~2b, right axis) \\
\hline
\end{tabular}
\end{table}

\paragraph{Monolayer MoTe$_2$.}
For MoTe$_2$ monolayers, a single, fully digitized exciton-resolved broadband absorptance dataset is not readily available in a machine-readable form suitable for direct open-source reproduction.
To enable fully reproducible regeneration of \textbf{Figure~4a} of Main text in an open-source workflow while retaining the essential near-$\Eg$ optical physics, we model the monolayer absorptance as a smooth ``continuum'' background plus a small set of excitonic resonances:
\begin{align}
a_{\mathrm{raw}}(E) &= a_{\mathrm{bg}}(E) + \sum_{j\in\{A,B,S\}} A_j\,L(E;E_j,\gamma_j)
\label{eq:mote2_araw}\\
a_{\mathrm{bg}}(E) &= A_{\mathrm{bg,sat}}\,S(E;E_{\mathrm{bg}},w_{\mathrm{bg}})
\label{eq:mote2_abg}\\
L(E;E_0,\gamma) &= \frac{\gamma^2}{(E-E_0)^2+\gamma^2}
\label{eq:lorentz_unit}\\
S(E;E_0,w) &= \tfrac{1}{2}\left[1+\tanh\!\left(\frac{E-E_0}{w}\right)\right]
\label{eq:smooth_step}
\end{align}
Here, $L(E;E_0,\gamma)$ is a unit-peak Lorentzian (so $A_j$ directly sets the peak height above background prior to global scaling), and $S(E;E_0,w)$ is a smooth step that approximates the onset of a weak continuum.
The resonances $j\in\{A,B\}$ represent the A/B excitons near the optical $\Eg$, while $j=S$ represents a weak higher-energy shoulder.

To set an absolute scale for the phenomenological MoTe$_2$ spectrum, we apply the same one-point A-exciton anchoring described above (Table~\ref{tab:mono_abs_anchor}).
We first compute $a_{\mathrm{raw}}(E)$ from Eqs.~(\ref{eq:mote2_araw})--(\ref{eq:smooth_step}), then scale it by a constant factor $s$ such that
the absorptance at the A-exciton peak energy $E_A$ matches the literature value $a_{\mathrm{ref}}(E_A)$:
\begin{align}
s &\equiv \frac{a_{\mathrm{ref}}(E_A)}{a_{\mathrm{raw}}(E_A)} \label{eq:mote2_scale}\\
a(E) &\equiv \min\!\left[a_{\mathrm{clip}},\, s\,a_{\mathrm{raw}}(E)\right]\label{eq:mote2_final}
\end{align}
For MoTe$_2$, the free-standing monolayer absorbance reported previously\cite{Ruppert2014MoTe2} gives $a_{\mathrm{ref}}(E_A)=0.026$ at $E_A=1.10$~eV.
We use a conservative numerical cap $a_{\mathrm{clip}}=0.12$ and obtain $s=0.30$ for our parameter set and energy grid. The complete parameter set is listed in Table~\ref{tab:mote2_mono_pheno}, while the calibrated absorption spectra used in \textbf{Figure~4a} of Main text are provided in the GitHub repository referenced in the main text.

\begin{table}[H]
\centering
\caption{Phenomenological parameters used to construct the monolayer MoTe$_2$ absorptance spectrum (Eqs.~\eqref{eq:mote2_araw}--\eqref{eq:mote2_final}). Values are listed before global scaling by $s$ [Eq.~\eqref{eq:mote2_scale}].}
\label{tab:mote2_mono_pheno}
\begin{tabular}{@{}lll p{0.46\textwidth}@{}}
\toprule
Component & Parameter & Value & Comment \\
\midrule
Optical gap anchor & $E_g$ & 1.10 eV & Optical gap anchor used for peak placement \\
Background step & $A_{\mathrm{bg,sat}}$ & 0.030 & Continuum onset (smooth step background) \\
 & $E_{\mathrm{bg}}$ & $E_g+0.18$ eV &  \\
 & $w_{\mathrm{bg}}$ & 0.35 eV &  \\
\addlinespace
A exciton (A) & $E_A$ & 1.10 eV & Band-edge exciton resonance \\
 & $\gamma_A$ & 0.030 eV &  \\
 & $A_A$ & 0.060 &  \\
\addlinespace
B exciton (B) & $E_B$ & 1.30 eV & Spin--orbit split exciton resonance \\
 & $\gamma_B$ & 0.060 eV &  \\
 & $A_B$ & 0.030 &  \\
\addlinespace
Shoulder (S) & $E_S$ & 1.90 eV & Higher-energy shoulder / continuum transitions \\
 & $\gamma_S$ & 0.200 eV &  \\
 & $A_S$ & 0.020 &  \\
\addlinespace
Anchor scaling & $a_{\mathrm{ref}}(E_A)$ & 0.026 & A-exciton absorbance anchor (Ref.~\cite{Ruppert2014MoTe2}, Fig.~2b) \\
 & $a_{\mathrm{clip}}$ & 0.12 & Numerical safety cap \\
\bottomrule
\end{tabular}
\end{table}

\paragraph{Validation and robustness: why a hard cutoff $a(E<E_g)=0$ is enforced.}
The phenomenological MoTe$_2$ monolayer model in Eqs.~\eqref{eq:mote2_araw}--\eqref{eq:mote2_final} contains Lorentzian resonances, which mathematically have nonzero tails at all energies.
If such tails are inserted directly into the Shockley-Queisser (SQ) and carrier multiplication (CM) DB integrals, the generalized Planck factor in Eq.~\eqref{eq:planck} becomes ill-defined whenever the photon chemical potential exceeds the photon energy ($\mu_\gamma>E$), because the Bose--Einstein denominator can approach zero or become negative.
In conventional DB, this pathology is avoided implicitly by the step absorptance $a(E)=\Theta(E-E_g)$, which guarantees that emissivity is strictly zero below $\Eg$, so that the physically required inequality $\mu_\gamma\le E$ is satisfied for all contributing energies.

In our phenomenological spectrum, the A-exciton Lorentzian centered at $E_A\approx E_g$ produces a substantial sub-$\Eg$ tail; as a result, maximizing $P(V)=JV$ in Eqs.~\eqref{eq:JSQ}--\eqref{eq:JCM} collapses the optimum voltage toward the smallest photon energies, where $a(E)$ is nonzero.
This effect is not a statement about real MoTe$_2$ device voltage; rather, it reflects a thermodynamic consistency constraint of the simplified line-shape model, when coupled to the black-body emission formula.
To preserve DB in a transparent open-source way, we, therefore, enforce a hard cutoff
\begin{equation}
a(E)=0\quad\text{for}\quad E<E_g
\end{equation}
in the generation script (\protect\path{build_absorptance_mote2_mono_pheno.py}) available in the GitHub repository referenced above.
Table~\ref{tab:mote2_cutoff_validation} quantifies the impact of this step: without the cutoff, both SQ and CM optimizations drive $V_{\mathrm{opt}}\rightarrow 0$ and the predicted efficiencies become artificially small; with the cutoff, $V_{\mathrm{opt}}$ returns to a physically reasonable fraction of $E_g/q$ and the efficiencies recover.

\begin{table}[H]
\centering
\caption{Validation of the hard-cutoff enforcement for the phenomenological MoTe$_2$ monolayer absorptance model. We compare SQ and CM (with $\eta_{\mathrm{CM}}=0.97$) optimal voltage $V_{\mathrm{opt}}$ and maximum efficiency $\eta$ computed using (i) the raw Lorentzian spectrum $a_{\mathrm{raw}}(E)$ [Eqs.~\eqref{eq:mote2_araw}--\eqref{eq:mote2_final} but without enforcing $a(E<E_g)=0$] and (ii) the thermodynamically consistent cutoff spectrum used throughout this work. The ``no cutoff'' case yields an unphysical voltage collapse, because $a(E)$ remains nonzero for $E<qV$.}
\label{tab:mote2_cutoff_validation}
\begin{tabular}{@{}lccc@{}}
\toprule
Case & Cutoff below $E_g$ & $V_{\mathrm{opt}}$ (V) & $\eta$ (\%) \\
\midrule
MoTe$_2$ mono, SQ & No  & 0.024 & 0.026 \\
MoTe$_2$ mono, SQ & Yes & 0.757 & 1.53 \\
MoTe$_2$ mono, CM & No  & 0.024 & 0.032 \\
MoTe$_2$ mono, CM & Yes & 0.760 & 1.71 \\
\bottomrule
\end{tabular}
\end{table}
\subsection{S1.3 CM quantum yield}
CM enters DB as an energy-dependent number of collected electron--hole pairs per absorbed photon, $m(E)\ge 1$.
We use a Beard-type piecewise-linear model
\begin{equation}
m(E)=
\begin{cases}
1, & E<2\Eg\\
1+\eta_{\mathrm{CM}}\left(\frac{E}{\Eg}-2\right), & 2\Eg\le E<3\Eg\\
2+\eta_{\mathrm{CM}}\left(\frac{E}{\Eg}-3\right), & 3\Eg\le E<4\Eg\\
\cdots &
\end{cases}
\label{eq:beard_m}
\end{equation}
with slope parameter $\eta_{\mathrm{CM}}$ used to match the previously measured quantum-yield ($\eta_{\mathrm{CM}}=0.97$).\cite{NatCommun2019CM}
We emphasize that this value is a deliberately optimistic upper bound derived from transient-absorption measurements under high-excitation conditions.
Under steady-state one-sun photovoltaic operation, actual CM yields may be lower.
Using this upper bound therefore stress-tests the absolute thermodynamic limit; any reduction in operational $\eta_{\mathrm{CM}}$ would further strengthen our conclusion that CM is negligible in wide-gap monolayer TMDs.

\section{S2. Generalized detailed balance (DB)}
\subsection{S2.1 SQ limit for non-ideal absorptance}
Kirchhoff's law equates emissivity to absorptance, so the radiative emission rate spectrum of a solar cell at temperature $\Tc$ and photon chemical potential $\mu_\gamma=\q V$ is proportional to $a(E)$.\cite{Rau2007Reciprocity,Kirchartz2018GoodCell}
The net extracted current density is
\begin{equation}
J_{\mathrm{SQ}}(V)=\q\int_0^\infty a(E)\left[\phi_{\odot}(E)-\phi_{\mathrm{bb}}(E,\Tc,\q V)\right]\dd E
\label{eq:JSQ}
\end{equation}
The electrical power density is $P(V)=J(V)V$, and the SQ efficiency is $\eta=\max_V P(V)/P_{\odot}$, where $P_{\odot}$ is the incident solar power density.
\subsection{S2.2 CM DB and the radiative penalty}
With CM, the collected carrier-pair yield per absorbed photon becomes $m(E)$.
The photogenerated current increases by $m(E)$, but detailed balance requires a corresponding increase in radiative recombination for a given terminal voltage because the photon chemical potential scales with the quasi-Fermi splitting that drives emission.\cite{Queisser1996CM,Beard2010CM}
Importantly, this generalized radiative term is not introduced ad hoc in the present work. Werner, Brendel, and Queisser explicitly noted that for solar cells with internal carrier multiplication the usual SQ assignment $\mu_\gamma=\q V$ for emitted photons is invalid and that the saturation current must be modified accordingly; Brendel et al. then formulated the corresponding thermodynamic efficiency limits, and later CM efficiency-limit studies by Klimov and by Hanna and Nozik continued to use detailed-balance models of this class.\cite{Werner1995APL,Queisser1996CM,Klimov2006APL,HannaNozik2006JAP}
A standard DB-consistent form, following Werner, Brendel, and Queisser and later used by Klimov and Hanna--Nozik, is
\begin{equation}
J_{\mathrm{CM}}(V)=\q\int_0^\infty m(E)\,a(E)\left[\phi_{\odot}(E)-\phi_{\mathrm{bb}}(E,\Tc,m(E)\q V)\right]\dd E
\label{eq:JCM}
\end{equation}
Equation~\eqref{eq:JCM} can be viewed as an internal-CM model: each absorbed photon produces $m(E)$ pairs, and the reverse radiative channel that annihilates those pairs carries the combined photon chemical potential $m(E)\q V$.\cite{Werner1995APL,Queisser1996CM,Klimov2006APL,HannaNozik2006JAP}
This should not be confused with assigning different electronic quasi-Fermi splittings to different carrier energies. The electronic system is still described by a single pair chemical potential; the factor $m(E)$ arises from the reaction stoichiometry of the inverse radiative channel.
As $m(E)$ increases, the radiative dark current increases faster, producing a voltage penalty that partially offsets the extra photocurrent.
\section{S3. Hot-carrier (HC) DB}
\subsection{S3.1 HC absorber as a nonequilibrium carrier reservoir}
In the HC model, absorbed photons thermalize among carriers to a hot electronic temperature $\Th$ before carriers cool to the lattice at $\Tc$.\cite{NatRevPhys2021HotCarrier}
The carrier population is described by a hot quasi-Fermi splitting $\muEH$ and temperature $\Th$.
Radiative emission is then governed by $\phi_{\mathrm{bb}}(E,\Th,\mu_\gamma=\muEH)$ (the photon chemical potential equals the e--h chemical potential under radiative equilibrium).
The HC current is therefore
\begin{equation}
J_{\mathrm{HC}}(\muEH,\Th)=\q\int_0^\infty a(E)\left[\phi_{\odot}(E)-\phi_{\mathrm{bb}}(E,\Th,\muEH)\right]\dd E
\label{eq:JHC}
\end{equation}
In the combined CM--HC model, an absorbed photon of energy $E$ produces an energy-dependent carrier yield $m(E)$, so the extracted pair flux scales with $m(E)$.
To preserve detailed balance, the reverse radiative process for photons in a given energy interval must annihilate $m(E)$ e--h pairs simultaneously (inverse Auger), which is captured by an energy-dependent photon chemical potential $\mu_\gamma(E)=m(E)\muEH$.
The CM--HC current is therefore
\begin{equation}
J_{\mathrm{CM\text{--}HC}}(\muEH,\Th)=\q\int_0^\infty a(E)\left[m(E)\phi_{\odot}(E)-\phi_{\mathrm{bb}}\!\left(E,\Th,\mu_\gamma=m(E)\muEH\right)\right]\dd E
\label{eq:JCMHC}
\end{equation}
which reduces to Eq.~(\ref{eq:JHC}) when $m(E)=1$.

\paragraph{Note on thermodynamic consistency.}
The use of an energy-dependent photon chemical potential $\mu_\gamma(E)=m(E)\muEH$ should be understood as the standard macroscopic detailed-balance extension for inverse-Auger emission, not as an explicit microscopic transport solution for an arbitrary nonequilibrium carrier distribution. The relevant endoreversible reaction for a radiative channel at energy $E$ is
\begin{equation}
m(E)\,(e\text{--}h) \rightleftharpoons \gamma_E,
\end{equation}
so the channel affinity vanishes when
\begin{equation}
\mathcal{A}_E = m(E)\muEH-\mu_\gamma(E)=0,
\end{equation}
which gives
\begin{equation}
\mu_\gamma(E)=m(E)\muEH.
\end{equation}
This derivation makes clear that the $E$ dependence belongs to the photon \emph{channel}, not to multiple electronic quasi-Fermi splittings. The hot electronic reservoir is still characterized by one pair chemical potential $\muEH$ and one carrier temperature $\Th$.

This is also the logic adopted in the foundational CM detailed-balance literature. Werner et al. first emphasized that the usual SQ assumption $\mu_\gamma=\q V$ is invalid for internal carrier multiplication; Brendel et al. then used the modified radiative balance to derive CM thermodynamic limits; and later efficiency-limit analyses by Klimov and by Hanna and Nozik continued to use detailed-balance CM formalisms of this type.\cite{Werner1995APL,Queisser1996CM,Klimov2006APL,HannaNozik2006JAP}

Accordingly, the present formulation should be read as an \emph{endoreversible upper-bound closure}. A fully microscopic device model would require explicit carrier-energy distributions, Auger kinetics, and phonon-cooling dynamics, but that is a different modeling objective from the thermodynamic benchmark developed here.
Within this endoreversible closure, the reversible limit $\muEH\to0$ forces the HC and CM--HC emission spectra to become identical, which is why the two limits are degenerate at $\kappa=0$.

\subsection{S3.2 Energy balance and the cooling-leakage term}
Define the absorbed and emitted energy flux densities
\begin{align}
P_{\mathrm{abs}} &= \int_0^\infty E\,a(E)\phi_{\odot}(E)\,\dd E\\
P_{\mathrm{emit}}(\muEH,\Th) &= \int_0^\infty E\,a(E)\phi_{\mathrm{bb}}(E,\Th,\muEH)\,\dd E
\end{align}
The net radiative energy deposited into the HC absorber is $P_{\mathrm{abs}}-P_{\mathrm{emit}}$.
We introduce a phenomenological heat-leakage term
\begin{equation}
\dot Q_{\mathrm{cool}}=\kappaC(\Th-\Tc)
\label{eq:cooling}
\end{equation}
representing phonon-mediated cooling (or any parasitic heat flow) that removes energy from the hot-carrier population.
If $\dot N = J/\q$ is the extracted carrier-pair flux, we define an average extracted carrier energy per pair $\Delta E$ via energy conservation,
\begin{equation}
\Delta E \equiv \frac{P_{\mathrm{abs}}-P_{\mathrm{emit}}-\dot Q_{\mathrm{cool}}}{\dot N}
\label{eq:deltaE_def}
\end{equation}
This $\Delta E$ is the energy window the energy-selective contacts must transmit (electron extraction at energy $E_e$ and hole extraction at $E_h$ such that $\Delta E=E_e-E_h$).

\paragraph{Physical interpretation of $\kappa$.}
The parameter $\kappa$ bridges the macroscopic heat-engine description and microscopic carrier cooling.
$\kappa=0$ is the perfectly reversible benchmark.
$\kappa=0.2~\si{\watt\per\meter\squared\per\kelvin}$ should be viewed as a highly optimistic aspirational target for engineered devices with strongly suppressed cooling.
For $\Delta T\approx500$~K, this corresponds to $\dot Q_{\mathrm{cool}}\approx100~\si{\watt\per\meter\squared}=10~\si{\milli\watt\per\centi\meter\squared}$, i.e., about 10\% of one-sun input power.
By contrast, $\kappa$ of order unity already implies several hundred \si{\watt\per\meter\squared} of heat leakage at comparable $\Delta T$, rapidly collapsing HC gains toward the SQ limit.
Accordingly, $\kappa$ is used here as a realizability parameter, not as a fitted microscopic material constant.

\subsection{S3.3 Energy-selective contacts and terminal voltage relation}
Energy-selective contacts convert part of $\Delta E$ into electrical work.
Under endoreversible extraction (reversible contacts; no entropy production at the contacts), the terminal voltage satisfies the De~Vos--Queisser relation\cite{DeVos1993,Sergeev2018}
\begin{equation}
\q V = \left(1-\frac{\Tc}{\Th}\right)\Delta E + \frac{\Tc}{\Th}\muEH.
\label{eq:devos_voltage}
\end{equation}
Equation~\eqref{eq:devos_voltage} can be also derived microscopically as described in Section~S4 (by equating the Fermi occupations at the selective-contact energies for hot and cold reservoirs).
The HC electrical power density is then $P_{\mathrm{HC}}=J_{\mathrm{HC}}V$, and the HC efficiency is obtained by maximizing $P_{\mathrm{HC}}$ over $(\muEH,\Th)$ (and, equivalently, over $V$ via Eq.~\eqref{eq:devos_voltage}).

\subsection{S3.4 Implementation details}
\label{sec:impl}
Our numerical procedure is:
(i) choose a material $(\Eg, a(E,d))$ and a CM model $m(E)$;
(ii) for SQ/CM compute $J(V)$ via Eqs.~\eqref{eq:JSQ}--\eqref{eq:JCM} and maximize $P(V)$;
(iii) for HC/CM--HC scan $(\muEH,\Th)$, compute $J$ and $P_{\mathrm{abs}}-P_{\mathrm{emit}}$ via Eqs.~\eqref{eq:JHC}--\eqref{eq:deltaE_def}, compute $V$ from Eq.~\eqref{eq:devos_voltage}, and maximize $P=JV$.
The public code package includes helper routines that write the numerical checks reported below: (i) a reversible HC/CM--HC degeneracy check, (ii) a cooling-budget table $\dot Q_{\mathrm{cool}}=\kappaC\Delta T$, (iii) a CM upper-bound stress test, and (iv) maximum-power current--voltage resource-accounting tables.

\subsection{S3.5 Reversible HC/CM--HC degeneracy and no double counting}
This subsection makes explicit the central thermodynamic resource-accounting point.  In the reversible hot-carrier limit ($\kappaC=0$), the optimum of the endoreversible HC problem is reached at $\muEH\rightarrow0$ for the optically thick bandgap range considered here.  At this point,
\begin{equation}
\mu_\gamma(E)=\muEH=0 \quad \mathrm{(HC)},
\end{equation}
whereas the combined CM--HC expression gives
\begin{equation}
\mu_\gamma(E)=m(E)\muEH=0 \quad \mathrm{(CM-HC)}.
\end{equation}
Thus, the emission spectrum is identical in HC and CM--HC at the reversible optimum.  The absorbed photon-energy flux is also identical because absorption energy is determined by the incident photons, not by the number of pairs into which the energy is partitioned.  Consequently, the reversible extracted power can be written as
\begin{equation}
P_{\mathrm{rev}}=\left(1-\frac{T_C}{T_H}\right)\left(P_{\mathrm{abs}}-P_{\mathrm{emit}}\right),
\end{equation}
which contains no independent CM contribution.  CM changes the particle flux $\dot N$ and therefore the voltage-per-pair scale, but it does not create an additional source of extractable work.

This degeneracy is a required consistency check.  A calculation in which HC and CM--HC gave different reversible ceilings under the same optical boundary conditions would double count the same excess photon-energy reservoir.  Finite-$\kappaC$ differences are allowed because the heat-leak term subtracts $\kappaC(T_H-T_C)$ from the hot reservoir and thereby changes the optimal allocation between voltage generation and current generation.

\subsection{S3.6 Scope of the upper-bound closure versus a predictive transport model}
The present model is deliberately not a microscopic kinetic simulation.  It assumes (i) rapid electronic thermalization to a hot carrier reservoir, (ii) radiative detailed balance with the generalized CM radiative penalty, (iii) ideal ESCs that realize the reversible De~Vos relation, and (iv) a lumped cooling conductance $\kappaC$ that represents all irreversible heat leakage from the hot reservoir.  A predictive device model would have to specify carrier-energy distributions, Auger rates, intervalley scattering, phonon bottlenecks, interface thermal conductance, nonradiative recombination, and finite-width contact transmission.  Those ingredients are material- and device-stack specific.  Our purpose is different: to provide a reproducible upper-bound benchmark that identifies where such device engineering could matter and where it cannot matter even under optimistic assumptions.

\section{S4. Microscopic flux derivation for energy-selective contacts (ESCs)}
\label{sec:microscopic}
This section provides the microscopic flux derivation that underpins Eq.~\eqref{eq:devos_voltage}.
\subsection{S4.1 Two-reservoir model and energy filtering}
Consider a hot electronic reservoir (the absorber) characterized by $(\Th,\muEH)$ coupled to two cold contacts (electron and hole reservoirs) at $(\Tc,\mu_e)$ and $(\Tc,\mu_h)$, where $\mu_e-\mu_h=\q V$.
ESCs transmit carriers only in narrow energy windows centered at $E_e$ (electron contact) and $E_h$ (hole contact), with $E_e>E_h$.
A generic Landauer-like form for the particle flux through an energy filter is
\begin{equation}
\dot N_e = \mathcal{T}_e(E_e)\left[f_{\mathrm{H}}(E_e;\Th,\mu_e^{\mathrm{H}})-f_{\mathrm{C}}(E_e;\Tc,\mu_e)\right]
\label{eq:landauer_e}
\end{equation}
where $f$ is the Fermi--Dirac occupation at the transmitted energy and $\mathcal{T}_e$ is a proportionality constant (transmission $\times$ density-of-states factor). An analogous expression holds for holes at $E_h$.
For an ideal infinitely narrow filter, only the occupations at $E_e$ and $E_h$ matter.
\subsection{S4.2 Reversible (endoreversible) contact condition}
A reversible contact implies zero entropy production at the interface, which is achieved, when there is no net driving force at the filter energy.
For the electron contact, reversibility is obtained, when the occupations match:
\begin{equation}
f_{\mathrm{H}}(E_e;\Th,\mu_e^{\mathrm{H}})=f_{\mathrm{C}}(E_e;\Tc,\mu_e)
\label{eq:match_e}
\end{equation}
Similarly for holes,
\begin{equation}
f_{\mathrm{H}}(E_h;\Th,\mu_h^{\mathrm{H}})=f_{\mathrm{C}}(E_h;\Tc,\mu_h)
\label{eq:match_h}
\end{equation}
The hot-reservoir electron and hole chemical potentials satisfy $\mu_e^{\mathrm{H}}-\mu_h^{\mathrm{H}}=\muEH$.
Using $f(E;T,\mu)=\left[1+\exp\!\left(\frac{E-\mu}{\kb T}\right)\right]^{-1}$, Eqs.~\eqref{eq:match_e}--\eqref{eq:match_h} imply
\begin{align}
\frac{E_e-\mu_e^{\mathrm{H}}}{\Th} &= \frac{E_e-\mu_e}{\Tc} \label{eq:linear_e}\\
\frac{E_h-\mu_h^{\mathrm{H}}}{\Th} &= \frac{E_h-\mu_h}{\Tc} \label{eq:linear_h}
\end{align}
Rearranging,
\begin{align}
\mu_e &= E_e-\frac{\Tc}{\Th}\left(E_e-\mu_e^{\mathrm{H}}\right)\\
\mu_h &= E_h-\frac{\Tc}{\Th}\left(E_h-\mu_h^{\mathrm{H}}\right)
\end{align}
Subtracting these two equations and using $\mu_e-\mu_h=\q V$, $\mu_e^{\mathrm{H}}-\mu_h^{\mathrm{H}}=\muEH$, and $\Delta E\equiv E_e-E_h$, we obtain
\begin{equation}
\q V=\left(1-\frac{\Tc}{\Th}\right)\Delta E + \frac{\Tc}{\Th}\muEH
\end{equation}
which is Eq.~\eqref{eq:devos_voltage}.
\subsection{S4.3 Heat and work per extracted pair}
The heat extracted from the hot reservoir per electron transmitted at energy $E_e$ is $Q_e=E_e-\mu_e^{\mathrm{H}}$; for a hole transmitted at $E_h$ it is $Q_h=\mu_h^{\mathrm{H}}-E_h$.
Thus, the total heat removed from the hot reservoir per extracted pair is
\begin{equation}
Q_{\mathrm{H}} = (E_e-\mu_e^{\mathrm{H}})+(\mu_h^{\mathrm{H}}-E_h)=\Delta E-\muEH
\end{equation}
The electrical work delivered to the external circuit per pair is $\q V$.
Combining with Eq.~\eqref{eq:devos_voltage} shows explicitly how a finite $\Th>\Tc$ enables conversion of part of the thermal energy $(\Delta E-\muEH)$ into electrical work, consistent with an endoreversible heat engine.

\section{S5. Additional numerical results and sensitivities}
\subsection{S5.1 MoTe$_2$ $\Eg$ sensitivity (optically thick absorber)}

\begin{figure*}[t]
\centering
\includegraphics[width=0.96\textwidth]{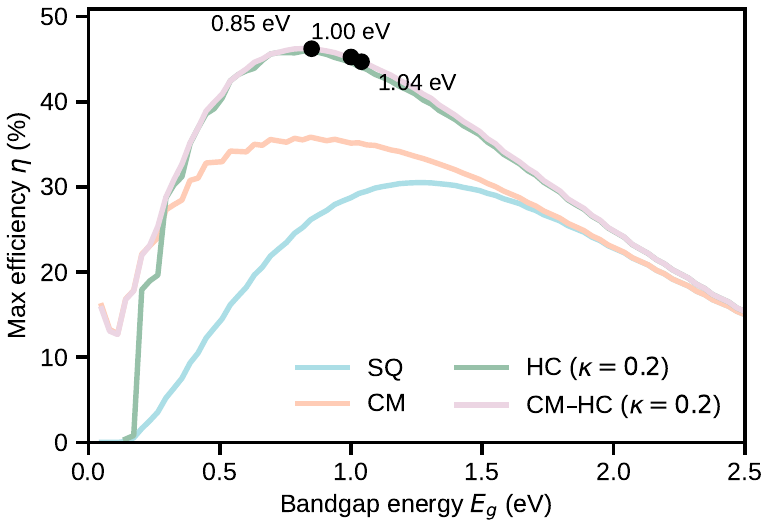}
\caption{$\Eg$ sweep for the optically thick absorber at $\kappaC=0.2$~\si{\watt\per\meter\squared\per\kelvin}, with marker points for the three MoTe$_2$ $\Eg$ used in Table~\ref{tab:MoTe2_main}.
This plot makes visually explicit why MoTe$_2$-like $\Eg$ cluster near the CM--HC optimum under \AM.}
\label{fig:MoTe2_markers}
\end{figure*}

Table~\ref{tab:MoTe2_main} summarizes the optically thick (step-function) results used to motivate MoTe$_2$ as a $\Eg$-optimal representative case for combined CM--HC operation. Figure~\ref{fig:MoTe2_markers} visualizes the same sensitivity with marker points at $\Eg=1.04$, 1.00, and 0.85~eV. 

\begin{table}[h]
\centering
\caption{MoTe$_2$ $\Eg$ sensitivity for an optically thick absorber under \AM\ using the Beard-type CM model with $\eta_{\mathrm{CM}}=0.97$ (main).
HC and CM--HC use $\kappaC=0.2$~\si{\watt\per\meter\squared\per\kelvin}.}
\label{tab:MoTe2_main}
\begin{tabular}{@{}llllll@{}}
\toprule
$\Eg$ (eV) & $\eta_{\mathrm{SQ}}$ (\%) & $\eta_{\mathrm{CM}}$ (\%) & $\eta_{\mathrm{HC}}$ (\%) & $\eta_{\mathrm{CM\mbox{--}HC}}$ (\%) & $\Delta\eta_{\mathrm{CM\mbox{--}HC}}$ (\%)   \\
\midrule
1.04 & 31.81 & 36.03 & 46.37 & 46.87 & 15.06   \\
1.00 & 31.54 & 36.44 & 47.19 & 47.66 & 16.12   \\
0.85 & 27.20 & 35.28 & 47.38 & 47.53 & 20.33   \\
\bottomrule
\end{tabular}
\end{table}

\subsection{S5.2 Thermodynamic consistency check: HC/CM--HC degeneracy at $\kappaC=0$}
Table~\ref{tab:degeneracy_check} reports the numerical check that HC and CM--HC become degenerate in the reversible limit.  The equality at $\kappaC=0$ is not a numerical accident; it is the resource-accounting result derived in Section~S3.5.  The same table also shows that finite positive CM--HC differences appear only when $\kappaC>0$.

\begin{table}[H]
\centering
\caption{Thermodynamic consistency check for representative bulk TMD bandgaps.  At $\kappaC=0$, HC and CM--HC have the same maximum efficiency to the precision reported.  At finite $\kappaC$, CM--HC can be slightly higher because CM repartitions part of the same above-gap energy into current before it is penalized by the heat leak.}
\label{tab:degeneracy_check}
\begin{tabular}{@{}lcccccc@{}}
\toprule
Material & $E_g$ (eV) & $\eta_{\mathrm{HC}}(0)$ & $\eta_{\mathrm{CM-HC}}(0)$ & $\Delta\eta(0)$ & $\eta_{\mathrm{HC}}(0.2)$ & $\eta_{\mathrm{CM-HC}}(0.2)$ \\
\midrule
WSe$_2$ & 1.29 & 47.6 & 47.6 & 0.0 & 39.7 & 40.1 \\
MoTe$_2$ & 1.04 & 51.4 & 51.4 & 0.0 & 44.1 & 44.8 \\
MoS$_2$ & 1.22 & 49.0 & 49.0 & 0.0 & 41.2 & 41.7 \\
\bottomrule
\end{tabular}
\end{table}

\subsection{S5.3 Maximum-power current--voltage resource accounting}
Table~\ref{tab:resource_accounting} makes the reversible current--voltage trade-off explicit.  HC and CM--HC have the same reversible efficiency, but CM--HC generally operates with a larger current density and lower voltage.  This is the numerical signature of repartitioning the same extracted power rather than accessing a new reversible thermodynamic channel.

\begin{table}[H]
\centering
\caption{Current--voltage resource accounting for bulk-like 50~nm absorbers at $\kappaC=0$.  The equal HC and CM--HC efficiencies are accompanied by higher $J_{\mathrm{mp}}$ and lower $V_{\mathrm{mp}}$ in CM--HC, demonstrating current--voltage repartitioning at fixed reversible power.}
\label{tab:resource_accounting}
\begin{tabular}{@{}lcccccc@{}}
\toprule
Material & $\eta_{\mathrm{HC}}$ & $J_{\mathrm{mp,HC}}$ & $V_{\mathrm{mp,HC}}$ & $\eta_{\mathrm{CM-HC}}$ & $J_{\mathrm{mp,CM-HC}}$ & $V_{\mathrm{mp,CM-HC}}$ \\
& (\%) & (mA cm$^{-2}$) & (V) & (\%) & (mA cm$^{-2}$) & (V) \\
\midrule
WSe$_2$ & 47.37 & 31.79 & 1.490 & 47.37 & 34.71 & 1.365 \\
MoTe$_2$ & 51.09 & 39.77 & 1.285 & 51.09 & 47.95 & 1.066 \\
MoS$_2$ & 48.66 & 33.90 & 1.435 & 48.66 & 37.81 & 1.287 \\
\bottomrule
\end{tabular}
\end{table}

\subsection{S5.4 Cooling-budget scale for interpreting $\kappaC$}
Table~\ref{tab:cooling_budget} converts the lumped cooling parameter into an areal heat leak using $\dot Q_{\mathrm{cool}}=\kappaC\Delta T$.  This table is included to avoid over-interpreting a single value of $\kappaC$.  Even the finite-leak benchmark used in the main text, $\kappaC=0.2~\si{\watt\per\meter\squared\per\kelvin}$, removes 60--160~W~m$^{-2}$ for $\Delta T=300$--800~K.  Thus, it is already an aspirational cooling-suppression target rather than a routine device condition.

\begin{table}[H]
\centering
\caption{Heat-leak budget $\dot Q_{\mathrm{cool}}=\kappaC\Delta T$ expressed as W~m$^{-2}$ and as a fraction of one-sun incident power (taken as 1000~W~m$^{-2}$).}
\label{tab:cooling_budget}
\begin{tabular}{@{}ccccccc@{}}
\toprule
$\kappaC$ & \multicolumn{2}{c}{$\Delta T=300$ K} & \multicolumn{2}{c}{$\Delta T=500$ K} & \multicolumn{2}{c}{$\Delta T=800$ K} \\
\cmidrule(lr){2-3}\cmidrule(lr){4-5}\cmidrule(lr){6-7}
(W m$^{-2}$ K$^{-1}$) & W m$^{-2}$ & 1-sun frac. & W m$^{-2}$ & 1-sun frac. & W m$^{-2}$ & 1-sun frac. \\
\midrule
0.00 & 0 & 0.00 & 0 & 0.00 & 0 & 0.00 \\
0.05 & 15 & 0.015 & 25 & 0.025 & 40 & 0.040 \\
0.20 & 60 & 0.060 & 100 & 0.100 & 160 & 0.160 \\
0.50 & 150 & 0.150 & 250 & 0.250 & 400 & 0.400 \\
1.00 & 300 & 0.300 & 500 & 0.500 & 800 & 0.800 \\
\bottomrule
\end{tabular}
\end{table}

\subsection{S5.5 Additional photovoltaic metrics ($J_\mathrm{sc}$, $V_\mathrm{oc}$, and FF)}

To facilitate a direct comparison beyond power conversion efficiency, Table~\ref{tab:pv_metrics_JV} summarizes the corresponding output metrics for the three representative TMD absorbers considered in \textbf{Figures~2--4}.

\begin{table}[H]
\centering
\caption{Photovoltaic metrics corresponding to the maximum-efficiency points reported for the three representative TMD absorbers in the main text. For the SQ and CM limits, $J_\mathrm{sc}$, $V_\mathrm{oc}$, and the fill factor (FF) are extracted from the diode-like $J(V)$ characteristic. For the endoreversible HC and CM--HC limits, the model optimizes over the energy-selective extraction parameter $\Delta E$ (Sec.~S4.5); because $\Delta E$ varies with operating point in this upper-bound framework, a unique full $J(V)$ curve (and hence $V_\mathrm{oc}$ and FF) is not defined without additional device-specific constraints. We therefore report the maximum-power-point metrics $(J_\mathrm{mp},V_\mathrm{mp})$ for HC and CM--HC, which directly correspond to the $\eta$ values plotted in Figs.~3--4.}
\label{tab:pv_metrics_JV}
\vspace{4pt}
\small
\setlength{\tabcolsep}{4.2pt}
\begin{tabular}{llcccccc}
\toprule
Material & Model & $\eta$ (\%) & $J_\mathrm{sc}$ & $V_\mathrm{oc}$ & FF & $J_\mathrm{mp}$ & $V_\mathrm{mp}$ \\
 &  &  & (mA~cm$^{-2}$) & (V) &  & (mA~cm$^{-2}$) & (V) \\
\midrule
\multicolumn{8}{l}{\textbf{Bulk-like films} (Yablonovitch limit, $d=50$~nm; $\kappa_\mathrm{C}=0$ for HC/CM--HC)}\\
\addlinespace
WSe$_2$ ($E_g=1.29$~eV) & SQ & 30.29 & 33.12 & 1.033 & 0.885 & 32.26 & 0.939 \\
 & CM & 33.03 & 36.03 & 1.036 & 0.885 & 35.10 & 0.941 \\
 & HC & 47.37 & \textemdash & \textemdash & \textemdash & 31.79 & 1.490 \\
 & CM--HC & 47.37 & \textemdash & \textemdash & \textemdash & 34.71 & 1.365 \\
\addlinespace
MoTe$_2$ ($E_g=1.04$~eV) & SQ & 29.08 & 42.21 & 0.801 & 0.861 & 40.75 & 0.714 \\
 & CM & 34.94 & 50.39 & 0.805 & 0.861 & 48.60 & 0.719 \\
 & HC & 51.09 & \textemdash & \textemdash & \textemdash & 39.77 & 1.285 \\
 & CM--HC & 51.09 & \textemdash & \textemdash & \textemdash & 47.95 & 1.066 \\
\addlinespace
MoS$_2$ ($E_g=1.22$~eV) & SQ & 30.25 & 35.53 & 0.968 & 0.879 & 34.54 & 0.876 \\
 & CM & 33.68 & 39.45 & 0.971 & 0.879 & 38.28 & 0.880 \\
 & HC & 48.66 & \textemdash & \textemdash & \textemdash & 33.90 & 1.435 \\
 & CM--HC & 48.66 & \textemdash & \textemdash & \textemdash & 37.81 & 1.287 \\
\addlinespace
\midrule
\multicolumn{8}{l}{\textbf{Monolayers} (anchored absorptance; $\kappa_\mathrm{C}=0$ for HC/CM--HC)}\\
\addlinespace
WSe$_2$ ($E_g=1.63$~eV) & SQ & 1.40 & 1.17 & 1.326 & 0.905 & 1.15 & 1.225 \\
 & CM & 1.43 & 1.19 & 1.327 & 0.906 & 1.17 & 1.227 \\
 & HC & 2.06 & \textemdash & \textemdash & \textemdash & 1.14 & 1.807 \\
 & CM--HC & 2.06 & \textemdash & \textemdash & \textemdash & 1.16 & 1.770 \\
\addlinespace
MoTe$_2$ ($E_g=1.10$~eV) & SQ & 0.46 & 0.62 & 0.846 & 0.866 & 0.60 & 0.757 \\
 & CM & 0.51 & 0.70 & 0.849 & 0.867 & 0.68 & 0.760 \\
 & HC & 0.76 & \textemdash & \textemdash & \textemdash & 0.59 & 1.293 \\
 & CM--HC & 0.76 & \textemdash & \textemdash & \textemdash & 0.66 & 1.149 \\
\addlinespace
MoS$_2$ ($E_g=1.89$~eV) & SQ & 1.52 & 1.07 & 1.560 & 0.917 & 1.05 & 1.458 \\
 & CM & 1.53 & 1.07 & 1.560 & 0.917 & 1.05 & 1.458 \\
 & HC & 2.13 & \textemdash & \textemdash & \textemdash & 1.04 & 2.048 \\
 & CM--HC & 2.13 & \textemdash & \textemdash & \textemdash & 1.04 & 2.038 \\
\addlinespace
\bottomrule
\end{tabular}
\end{table}

\subsection{S5.6 $\kappaC$ sweep}
Table~\ref{tab:kappa_values} gives the numerical values underlying the $\kappaC$-sweep figure used in the main-text discussion of cooling sensitivity.
The sweep values $\kappaC=\{0,0.05,0.2,0.5,1.0\}~\si{\watt\per\meter\squared\per\kelvin}$ span 1--2 decades and are convenient for visualizing how rapidly HC gain collapses with cooling leakage.
\begin{table}[H]
\centering
\caption{Numerical values underlying \textbf{Figure~2(b)} of Main text, reported as absolute efficiencies (\%) at the bulk $\Eg$ $E_g=1.29$~eV (WSe$_2$), $1.04$~eV (MoTe$_2$), and $1.22$~eV (MoS$_2$).}
\label{tab:kappa_values}
\begin{tabular}{@{}ccccccc@{}}
\toprule
& \multicolumn{2}{c}{WSe$_2$} & \multicolumn{2}{c}{MoTe$_2$} & \multicolumn{2}{c}{MoS$_2$} \\
\cmidrule(lr){2-3} \cmidrule(lr){4-5} \cmidrule(lr){6-7}
$\kappaC$ (W m$^{-2}$ K$^{-1}$) & HC & CM--HC & HC & CM--HC & HC & CM--HC   \\
\midrule
0.00 & 47.6 & 47.6 & 51.4 & 51.4 & 49.0 & 49.0   \\
0.05 & 45.1 & 45.1 & 49.5 & 49.5 & 46.6 & 46.6   \\
0.20 & 39.7 & 40.1 & 44.1 & 44.8 & 41.2 & 41.7   \\
0.50 & 36.3 & 37.3 & 39.0 & 41.2 & 37.3 & 38.7   \\
1.00 & 34.2 & 36.0 & 35.7 & 38.9 & 34.8 & 37.0   \\
\bottomrule
\end{tabular}
\end{table}

\subsection{S5.7 CM upper-bound stress test at $\eta_{\mathrm{CM}}=0.97$}
Table~\ref{tab:cm_stress_test} isolates the pure CM effect by comparing SQ and CM results at the intentionally optimistic quantum-yield ceiling.  This is a stress test in favor of CM.  The wide-gap monolayers still show only tiny absolute gains, whereas the narrow-gap, optically stronger bulk-like systems show a larger CM contribution.

\begin{table}[H]
\centering
\caption{CM upper-bound stress test.  Values are from the same maximum-efficiency points reported in Table~S5.  Relative gains are computed from the SQ and CM entries.}
\label{tab:cm_stress_test}
\begin{tabular}{@{}lccccc@{}}
\toprule
Absorber & $E_g$ (eV) & $\eta_{\mathrm{SQ}}$ (\%) & $\eta_{\mathrm{CM}}$ (\%) & $\Delta\eta_{\mathrm{CM}}$ (\%) & $\Delta J_{\mathrm{sc}}/J_{\mathrm{sc}}$ (\%) \\
\midrule
Bulk WSe$_2$ (50 nm) & 1.29 & 30.29 & 33.03 & 2.74 & 8.8 \\
Bulk MoTe$_2$ (50 nm) & 1.04 & 29.08 & 34.94 & 5.86 & 19.4 \\
Bulk MoS$_2$ (50 nm) & 1.22 & 30.25 & 33.68 & 3.43 & 11.0 \\
Monolayer WSe$_2$ & 1.63 & 1.40 & 1.43 & 0.03 & 1.7 \\
Monolayer MoTe$_2$ & 1.10 & 0.46 & 0.51 & 0.05 & 12.9 \\
Monolayer MoS$_2$ & 1.89 & 1.52 & 1.53 & 0.01 & 0.0 \\
\bottomrule
\end{tabular}
\end{table}

\subsection{S5.8 Traceability of added calculations}
The additional tables above are generated from the same detailed-balance functions used for the main figures plus simple post-processing scripts.  The code package contains named outputs for Tables~\ref{tab:degeneracy_check}--\ref{tab:resource_accounting}; these files can be compared directly with the values shown here.

\clearpage
\FloatBarrier
\bibliography{refs}